\documentstyle[11pt,amsmath,amssymb]{article}

\numberwithin{equation}{section}

\newcommand{\Ll}{L_{\lambda_1,\lambda_2}}
\newcommand{\Kl}{K_{\lambda_1,\lambda_2}}
\newcommand{\Pbar}{{\overline P}}
\newcommand{\Lbar}{{\overline L}}
\newcommand{\fd}{{\frak D}}
\newcommand{\vb}{\underline{\varrho}}
\newcommand{\rhobar}{\overline{\rho}}
\newcommand{\beq}{\begin{equation}}
\newcommand{\eeq}{\end{equation}}
\newcommand{\bex}{\begin{example}}
\newcommand{\eex}{\end{example}}
\newcommand{\ber}{\begin{remark}}
\newcommand{\eer}{\end{remark}}
\def\bel{\begin{lemma}}
\def\eel{\end{lemma}}
\def\bet{\begin{theorem}}
\def\eet{\end{theorem}}
\def\bed{\begin{definition}}
\def\eed{\end{definition}}
\def\bea{\begin{assumption}}
\def\eea{\end{assumption}}
\newcommand{\cc}{{\cal C}}

\newcommand{\bbbone}{\mathchoice {\rm 1\mskip-4mu l} {\rm 1\mskip-4mu l}
{\rm 1\mskip-4.5mu l} {\rm 1\mskip-5mu l}}
\newcommand{\scalprod}[2]{\left\langle {#1}, {#2}\right\rangle}

\newcommand{\IM}{{\rm Im}}
\newcommand{\fer}[1]{(\ref{#1})}

\newcommand{\h}{{\cal H}}
\newcommand{\cx}{{\mathbb C}}
\newcommand{\r}{{\mathbb R}}

\newcommand{\av}[1]{\left\langle{#1}\right\rangle}

\newcommand{\mm}{{\frak M}}

\newcommand{\spec}{{\rm spec}}
\newcommand{\e}{{\rm e}}

\newcommand{\s}{{\rm S}}

\newcommand{\rr}{{\rm R}}

\renewcommand{\i}{{\rm i}}

\renewcommand{\d}{{\rm d}}

\newcounter{resultcounter}[section]

\newtheorem{theorem}[resultcounter]{Theorem}
\newtheorem{lemma}[resultcounter]{Lemma}
\newtheorem{proposition}[resultcounter]{Proposition}

\newcommand{\usigma}{{\underline{\,\sigma\!}\,}}
\newcommand{\utau}{{\underline{\,\tau\!}\,}}

\begin{document}

\setcounter{page}{0}

\title{Dynamics of Collective Decoherence and Thermalization}

\author{
M. Merkli\footnote{ Department of Mathematics and Statistics,
Memorial University of Newfoundland, St. John's, NL, Canada A1C
5S7; Supported by NSERC under grant 205247; Email:
merkli@math.mun.ca; URL: http://www.math.mun.ca/\
$\widetilde{}$\,merkli/}\ \ \ G. P. Berman\footnote{Theoretical
Division and CNLS, MS B213, Los Alamos National Laboratory, Los
Alamos, NM 87545, USA; Supported by the NNSA of the U.S. DOE at
LANL under Contract No. DE-AC52-06NA25396; Email: gpb@lanl.gov} \
\ \ I. M. Sigal\footnote{Department of Mathematics,
University of Toronto, Toronto, ON, Canada M5S 2E4; Supported by
NSERC under grant NA 7901; Email: im.sigal@utoronto.ca; \ \ \ \ \
\ \ \ \ \ \   URL: http://www.math.toronto.edu/sigal/ } }
\date{\today}

\maketitle

\begin{abstract}
We analyze the dynamics of $N$ interacting spins (quantum
register) collectively coupled to a thermal environment. Each spin
experiences the same environment interaction, consisting of an
energy conserving and an energy exchange part.

We find the decay rates of the reduced density matrix elements in
the energy basis. We show that if the spins do not interact among each other,
then the fastest decay rates of off-diagonal matrix
elements induced by the energy conserving interaction is of order
$N^2$, while that one induced by the energy exchange interaction
is of the order $N$ only. Moreover,  the diagonal matrix
elements approach their limiting values at a rate independent of
$N$.
For a general spin system the decay rates depend
in a rather complicated (but explicit) way on the size $N$ and the interaction between the spins.

Our method is based on a dynamical quantum resonance theory valid
for small, fixed values of the couplings. We do not make Markov-,
Born- or weak coupling (van Hove) approximations.

\end{abstract}

\thispagestyle{empty}
\setcounter{page}{1}
\setcounter{section}{1}
\setcounter{section}{0}

\section{Introduction}
\label{sect1}

{\bf Description of the problem. } We consider a qubit register of
size $N$ whose Hamiltonian is of the form
\begin{equation}
H_\s = \sum_{i,j=1}^N J_{ij}S^z_i S^z_j +\sum_{j=1}^N B_j S_j^z,
\label{2}
\end{equation}
where the $J_{ij}$ are pair interaction constants that can take
positive or negative values, and $B_j\geq 0$ is an effective
magnetic field at the location of spin $j$
($B_j=\frac{\hbar}{2}\gamma B_j^z$, where $\hbar$ is the Planck
constant, $\gamma$ is the value of the electron gyromagnetic ratio
and $B_j^z$ is an inhomogeneous magnetic field, oriented in the
positive $z$ direction). Also,
\begin{equation}
S^z= \left[
\begin{array}{cc}
1 & 0\\
0 & -1
\end{array}
\right]
\label{4}
\end{equation}
is the Pauli spin $1/2$ operator; $S_j^z$ is the matrix $S^z$
acting nontrivially only on the $j$-th spin. The environment $\rr$
is modelled by a bosonic thermal reservoir  whose Hamiltonian is
\begin{equation}
H_\rr =\int_{\r^3} a^*(k) |k| a(k) \d^3k,
\label{5}
\end{equation}
where $a^*(k)$ and $a(k)$ are the usual bosonic creation and
annihilation operators satisfying the canonical commutation
relations $[a(k),a^*(l)]=\delta(k-l)$. It is understood that we consider $\rr$ in the
thermodynamic limit of infinite volume, fixed temperature
$T=1/\beta>0$, in a phase without Bose-Einstein condensate.

We consider a {\it collective coupling}: the distance between the
$N$ qubits is smaller than the correlation length of the reservoir
and consequently each qubit feels {\it the same} interaction with
the latter. The collective interaction between $\s$ and $\rr$ is
given by the operator
\begin{equation}
v=\lambda_1 v_1+\lambda_2 v_2 = \lambda_1\sum_{j=1}^N S_j^z\otimes
\phi(g_1) + \lambda_2\sum_{j=1}^N S_j^x\otimes\phi(g_2).
 \label{6}
\end{equation}
Here, $\phi(g)$ is the field operator smoothed out with a {\it
form factor} (coupling function) $g=g(k)$, $k\in{\mathbb R}^3$,
see \fer{7} in Appendix \ref{formulas}. The coupling constants
$\lambda_1$ and $\lambda_2$ measure the strengths of the {\it
energy conserving} (position-position) coupling, and the {\it
energy exchange} (spin flip) coupling, respectively. Spin-flips
are implemented by the $S_j^x$ in \fer{6}, representing the Pauli
matrix
\begin{equation}
S^x =\ \left[
\begin{array}{cc}
0 & 1\\
1 & 0
\end{array}
\right]
 \label{8}
\end{equation}
acting on the $j$-th factor of $\h_\s$.
The total Hamiltonian takes the form
\begin{equation}
H=H_\s+H_\rr +v.
 \label{9}
\end{equation}
The dynamics of a density matrix $\rho_t$ of the system $\s+\rr$
is governed by the Liouville-von Neumann equation
$$
\frac{\d}{\d t}\rho_t = -\i[H,\rho_t],
$$
with initial condition $\rho_t|_{t=0}=\rho_0$. The solution to the
Liouville-von Neumann equation is given by $\rho_t=\e^{-\i
tH}\rho_0\e^{\i tH}$. We are interested only in information on the
subsystem $\s$, so we trace out the degrees of freedom of $\rr$.
The state of $\s$ is given by the reduced density matrix
\begin{equation}
\rhobar_t = {\rm Tr}_{\rr}(\e^{-\i t H}\rho_0\e^{\i tH}),
\label{10}
\end{equation}
where $\rho_0$ is the initial density matrix of the coupled
system, and ${\rm Tr}_\rr$ is the partial trace over the degrees
of freedom of the reservoir. The operator $\rhobar_t$ acts on the
Hilbert space $\h_\s=\cx^2\otimes\cdots \otimes\cx^2 = \cx^{2^N}$
of $\s$ only.

 Our goal is to analyze the time evolution of matrix
elements of the reduced density matrix \fer{10} in the energy
basis, which plays a special  role in quantum information theory.
The energy basis consists of eigenvectors $\varphi_\usigma$ of
$H_\s$, indexed by spin configurations
\begin{equation}
\usigma=\{\sigma_1,\ldots,\sigma_N\}\in \{+1,-1\}^N,\qquad
\varphi_\usigma =
\varphi_{\sigma_1}\otimes\cdots\otimes\varphi_{\sigma_N}.
\label{11}
\end{equation}
Here,
\begin{equation}
\varphi_+=\left[
\begin{array}{c}
1\\
0
\end{array}
\right],\quad
 \varphi_-=\left[
\begin{array}{c}
0\\
1
\end{array}
\right],
\label{12}
\end{equation}
so that
\begin{equation}
H_\s\varphi_\usigma = E(\usigma)\varphi_\usigma \mbox{\qquad with
\qquad} E(\usigma) = \sum_{i,j=1}^N J_{ij}\sigma_i\sigma_j
+\sum_{j=1}^N B_j\sigma_j.
\label{13}
\end{equation}
 We denote the reduced
density matrix elements as
\begin{equation}
[\rhobar_t]_{\usigma,\utau} =
\scalprod{\varphi_\usigma}{\rhobar_t\varphi_\utau}. \label{14}
\end{equation}
The dynamics of the register alone (without coupling to the
environment) is given by $\rhobar_t=\e^{-\i t H_\s}\rhobar_0\e^{\i
t H_\s}$, where $\rhobar_0={\rm Tr}_\rr(\rho_0)$, so matrix
elements of $\rhobar_t$ have the time dependence
\begin{equation}
[\rhobar_t]_{\usigma,\utau} = \e^{\i
t\{E(\utau)-E(\usigma)\}}[\rhobar_0]_{\usigma,\utau}.
 \label{15}
\end{equation}
We view the energy differences
\begin{equation}
e(\usigma,\utau):= E(\usigma)-E(\utau)= \sum_{i,j=1}^N
J_{ij}(\sigma_i\sigma_j-\tau_i\tau_j) +\sum_{j=1}^N
B_j(\sigma_j-\tau_j)
 \label{e}
\end{equation}
as being eigenvalues of the {\it Liouville operator}
\begin{equation}
L_\s = H_\s\otimes\bbbone -\bbbone\otimes H_\s, \label{16}
\end{equation}
acting on the doubled space
\begin{equation}
\h_\s\otimes\h_\s = (\cx^2\otimes\cx^2)\otimes\cdots \otimes
(\cx^2\otimes\cx^2),
 \label{17}
\end{equation}
where the $j$-th pair  $\cx^2\otimes\cx^2$ is the doubled space of
the $j$-th qubit.

\bigskip
{\bf Discussion of main results.\ }
 In the {\it resonance approach} used in this work, we examine the influence of
the interaction \fer{6} on the free dynamics \fer{15} for small
coupling parameters $\lambda_1$, $\lambda_2$. Under the
perturbation, the phase factors $e=E(\utau)-E(\usigma)$ in
\fer{15} become {\it complex resonance energies}, $\varepsilon_e=
\varepsilon_e(\lambda_1,\lambda_2)\in\cx$. The latter encode
properties of irreversibility of the reduced dynamics of $\s$
(decay of observables and matrix elements -- the dynamics of the
entire system $\s+\rr$ is unitary, by contrast). We consider the
regime where the resonance energies
$\varepsilon_e(\lambda_1,\lambda_2)$ do not overlap as the
perturbation is switched on, so that each resonance energy can be
followed separately. This means that the coupling parameters must
be small with respect to the gap between differences of energies
of $H_\s$, see condition (A1) in Section \ref{resultsection}
below.\footnote{Our method is applicable as well if this condition
is not imposed. Work on this is in progress.} We make as well a
technical assumption (A2) on the regularity of form factors $g_1$
and $g_2$ which we explain in Section \ref{resultsection}.

\smallskip
{\bf Dynamics of $\s$.\ } Our first result is a detailed
description of the evolution of the reduced density matrix
elements (and hence of all observables). Set
\begin{equation}
\langle\!\langle[\rhobar_\infty]_{\usigma,\utau} \rangle\!\rangle
=\lim_{T\rightarrow\infty} \frac 1T \int_0^T
[\rhobar_t]_{\usigma,\utau} \ \d t.
 \label{2.9}
\end{equation}
We show in Theorem \ref{thm1} that this limit exists, and that for
all $t\geq 0$,
\begin{eqnarray}
[\rhobar_t]_{\usigma,\utau} - \langle\!\langle[
\rhobar_\infty]_{\usigma,\utau} \rangle\!\rangle &=& \sum_{\{e,s:\
\varepsilon_e^{(s)}\neq 0\}} \e^{\i t\varepsilon_e^{(s)}} \Big[
{\sum_{\usigma',\utau'}}^*\
w^{\varepsilon_e^{(s)}}_{\usigma,\utau; \usigma',\utau'}\
[\rhobar_0]_{\usigma',\utau'}
+O(\lambda_1^2+\lambda_2^2)\Big]\nonumber \\
&&  \quad + O\big((\lambda_1^2+\lambda_2^2)\e^{-\omega't}\big),
\label{intro1}
\end{eqnarray}
where ${\rm Im}\varepsilon_e^{(s)}\geq 0$ and $\omega'$ satisfies
$2 \max\{{\rm Im}\varepsilon_e^{(s)}\}\leq \omega'<\tau$, with
$\tau>0$ a constant depending on the regularity of  $g_1$, $g_2$
(see Condition (A2) in Section \ref{resultsection}, and also
\cite{MSB1}). The $*$ in the sum \fer{intro1} means that we sum
only over configurations $\usigma', \utau'$ such that
$e(\usigma',\utau')=-e$. The
coefficents $w$ are overlaps of resonance eigenstates (see Section
\ref{mres1}), which vanish unless $e(\usigma,\utau)=-e$, in which case they are $O(1)$ in $\lambda_1, \lambda_2$. The
$\varepsilon_e^{(s)}$ are eigenvalues of a certain explicit
operator $K(\omega')$, a ``spectrally deformed Liouville
operator'' (see Section \ref{sect3}). They have the expansion
\begin{equation}
\varepsilon^{(s)}_e = e +\delta_e^{(s)}
+O(\lambda_1^4+\lambda_2^4), \label{2.2}
\end{equation}
where the label $s=1,\ldots,\nu(e)$ indexes the splitting of the
eigenvalue $e$ of $L_\s$, having multiplicity $d(e)$, into
$\nu(e)\leq d(e)$ distinct resonance energies. The lowest order
corrections $\delta_e^{(s)}$ satisfy
\begin{equation}
\delta_e^{(s)}=O(\lambda_1^2+\lambda_2^2 ).
\label{intro2}
\end{equation}
They are the (complex) eigenvalues of an operator $\Lambda_e$,
called the {\it level shift operator} associated to $e$
($\Lambda_e$ is related to the Lindblad generator). This operator
acts on the eigenspace of $L_\s$ associated to the eigenvalue $e$
(a subspace of the qubit register Hilbert space; see equation
\fer{5.1} for the formal definition of $\Lambda_e$). It governs
the lowest order shift of eigenvalues under perturbation. One can
see by direct calculation that $\IM\,\delta_e^{(s)}\geq
0$.\footnote{This can also be inferred from general considerations
\cite{MMS}: If the imaginary part was negative, then the average
of some observables would explode as time increases, contradicting
the fact that the total dynamics, a group of automorphisms, cannot
increase indefinitely the average of any observable.}

{\bf Discussion of \fer{intro1}.\ } To lowest order in the
perturbation, the group of reduced density matrix elements
$[\rhobar_t]_{\usigma,\utau}$ associated to a fixed
$e=e(\usigma,\utau)$ evolve in a coupled way, while groups of
matrix elements associated to different $e$ evolve independently.
The density matrix elements of a given group mix and evolve in
time according to the weight functions $w$ and the exponentials
$\e^{\i t\varepsilon_e^{(s)}}$. In the absence of interaction
($\lambda_1=\lambda_2=0$) all the $\varepsilon_e^{(s)}$ are real.
As the interaction is switched on, the $\varepsilon_e^{(s)}$
typically migrate into the upper complex plane, but they may stay
on the real line in certain cases. The matrix elements
$[\rhobar_t]_{\usigma,\utau}$ of a group $e$ approach their
ergodic means \fer{2.9} if and only if all the nonzero
$\varepsilon_e^{(s)}$ have strictly positive imaginary part. In
this case the convergence takes place on a time scale of the order
$1/\gamma_e$, where
\begin{equation}
\gamma_e=\min\left\{
\IM\,\varepsilon_e^{(s)}:\
s=1,\ldots,\nu(e) {\ \rm s.t.\ } \varepsilon_e^{(s)}\neq 0\right\}
 \label{3.17.1}
\end{equation}
is the decay rate of the group associated to $e$. If an $\varepsilon_e^{(s)}$ stays real then the matrix elements of the coresponding group oscillate in time. A sufficient condition for decay of the group associated to $e$ is $\gamma_e>0$, i.e. ${\rm Im}\delta_e^{(s)}>0$ for all $s$, and $\lambda_1$, $\lambda_2$ small.


\smallskip
{\bf Decoherence rates.\ } We illustrate our results on
decoherence rates for a qubit register with $J_{ij}=0$ (the
general case is treated in Section \ref{sect02}).  We consider
{\it generic} magnetic fields defined as follows. For
$n_j\in\{0,\pm 1,\pm2\}$, $j=1,\ldots,N$, we have
\begin{equation}
\sum_{j=1}^N B_jn_j=0\qquad \Longleftrightarrow \qquad n_j=0\
\forall j. \label{3.1}
\end{equation}
Condition \fer{3.1} is satisfied generically in the sense that
only for very special choices of $B_j$ does it not hold (one such
special choice is $B_j={\mbox constant}$). For instance, if the
$B_j$ are chosen independent, and uniformly random from an
interval $[B_{\rm min},B_{\rm max}]$, then \fer{3.1} is satisfied
with probability one. We show in Theorem \ref{mlemmaX} that the
decoherence rates \fer{3.17.1} are given by
\begin{equation}
\gamma_e =
\left\{
\begin{array}{ll}
\lambda_1^2 y_1(e)+\lambda_2^2 y_2(e) +y_{12}(e)
, & e\neq 0\\
\lambda_2^2 y_0, & e=0
\end{array}
\right\}
+ O(\lambda_1^4+\lambda_2^4).
\label{3.22.1}
\end{equation}
Here, $y_1$ is a contributions coming from the energy conserving interaction, $y_0$ and $y_2$ are due to the spin flip interaction. The term $y_{12}$ is due to both interactions and is of $O(\lambda^2_1+\lambda^2_2)$. We give explicit expressions for $y_0$, $y_1$, $y_2$ and $y_{12}$ in equations \fer{fgr2}, \fer{y1}, \fer{3.11.1} and \fer{y12}.

\begin{itemize}

\item[-] {\it Properties of $y_1(e)$:} $y_1(e)$ vanishes if either
$e$ is such that $e_0:=\sum_{j=1}^n(\sigma_j-\tau_j)=0$, or the
infra-red behaviour of the coupling function $g_1$ is too regular
(in three dimensions $g_1\propto |k|^p$ with $p>-1/2$). Otherwise
$y_1(e)>0$. Moreover, $y_1(e)$ is proportional to the temperature
$T$.

\item[-] {\it Properties of $y_2(e)$:} $y_2(e)>0$ if $g_2(2B_j,\Sigma)\neq 0$ for all $B_j$ (form factor $g_2(k)=g_2(|k|,\Sigma)$ in spherical coordinates). For low temperatures $T$, $y_2(e)\propto T$, for high temperatures $y_2(e)$ approaches a constant.

\item[-] {\it Properties of $y_{12}(e)$:} If either of $\lambda_1$, $\lambda_2$ or $e_0$ vanish, or if $g_1$ is infra-red regular as mentioned above, then $y_{12}(e)= 0$. Otherwise $y_{12}(e)>0$, in which case $y_{12}(e)$ approaches constant values for both $T\rightarrow 0, \infty$.

\item[-] {\it Full decoherence:} If $\gamma_e>0$ for all $e\neq 0$
then all off-diagonal matrix elements approach their limiting
values exponentially fast. In this case we say that full
decoherence occurs. It follows from the above points that we have
full decoherence if $\lambda_2\neq 0$ and $g_2(2B_j,\Sigma)\neq 0$
for all $j$, and provided $\lambda_1,\lambda_2$ are small enough
(so that the remainder term in \fer{3.22.1} is small). Note that
if $\lambda_2=0$ then matrix elements associated to energy
differences $e$ such that $e_0=0$ will not decay on the time scale
given by the second order in the perturbation ($\lambda_1^2$). \\
We point out that generically, $\s+\rr$ will reach a joint equilibrium as $t\rightarrow\infty$, which means that the final reduced density matrix of $\s$ is its Gibbs state modulo a peturbation of the order of the interaction between $\s$ and $\rr$, see \cite{MSB1}. Hence generically, the density matrix of $\s$ does not become diagonal in the energy basis as $t\rightarrow\infty$.

\item[-] {\it Properties of $y_0$:} $y_0$ depends on the energy
exchange interaction only. This reflects the fact that for a
purely energy conserving interaction, the populations are
conserved \cite{MSB1,PSE}. If $g_2(2B_j,\Sigma)\neq 0$ for all
$j$, then $y_0>0$ (this is sometimes called the ``Fermi Golden
Rule Condition''). For small temperatures $T$, $y_0\propto T$,
while $y_0$ approaches a finite limit as $T\rightarrow\infty$.
\end{itemize}

In terms of complexity analysis, it is important to discuss the {\it dependence of $\gamma_e$ on the register size $N$}.

\begin{itemize}
\item[-] We see from \fer{fgr2} that $y_0$ is independent of $N$. This means that the thermalization time, or relaxation time of the diagonal matrix elements (corresponding to $e=0$), is $O(1)$ in $N$.

\item[-] To determine the order of magnitude of the
decay rates of the off-diagonal density matrix elements (corresponding to $e\neq 0$) relative to the register size $N$, we
assume the magnetic field to have a certain distribution denoted
by $\av{\ }$. It follows from the explicit expressions for $y_1$, $y_2$ and $y_{12}$ (see \fer{y1}, \fer{3.11.1} and \fer{y12}) that
\begin{equation}
\av{y_1}=y_1\propto e_0^2,\quad  \av{y_2}=C_B{\frak D}(\usigma-\utau), \quad \mbox{and}\quad \av{y_{12}}= c_B(\lambda_1,\lambda_2) N_0(e),
\label{3.20'}
\end{equation}
where $C_B$ and $c_B=c_B(\lambda_1,\lambda_2)$ are positive constants (independent of $N$), with $c_B(\lambda_1,\lambda_2)= O(\lambda_1^2+\lambda_2^2)$. Here, $N_0(e)$ is the number of indices $j$ such that $\sigma_j=\tau_j$ for each $(\usigma,\utau)$ s.t. $e(\usigma,\utau)=e$, and
\begin{equation}
\fd (\usigma-\utau):= \sum_{j=1}^N |\sigma_j-\tau_j|
 \label{3.20}
\end{equation}
is the {\it Hamming distance} between the spin configurations
$\usigma$ and $\utau$ (which depends on $e$ only).

\item[-] Consider $e\neq 0$. It follows from \fer{3.22.1}-\fer{3.20} that for purely energy conserving interactions ($\lambda_2=0$), $\gamma_e\propto \lambda_1^2 e_0^2=\lambda_1^2 [\sum_{j=1}^N(\sigma_j-\tau_j)]^2$, which can be as large as $O(\lambda_1^2 N^2)$. On the other hand, for purely energy exchanging interactions ($\lambda_1=0$), we have $\gamma_e\propto \lambda_2^2 \fd(\usigma-\utau)$, which cannot exceed $O(\lambda_2^2 N)$. If both interactions are acting, then we have the additional term $\av{y_{12}}$, which is of order $O((\lambda_1^2+\lambda_2^2)N)$. This shows the following:

{\it The fastest decay rate of reduced off-diagonal density matrix elements due to
the energy conserving interaction alone is of order $\lambda_1^2 N^2$,
while the fastest decay rate due to the energy exchange interaction alone is of the order $\lambda_2^2 N$. Moreover, the decay of the diagonal matrix elements is of oder $\lambda_1^2$, i.e., independent of $N$.}

\item[-] The same discussion is valid for the interacting register ($J_{ij}\neq 0$), see Section \ref{sect02}.

\end{itemize}

\noindent
{\bf Remarks.\ }
1. For $\lambda_2=0$ the model can be
solved explicitly \cite{PSE}, and one shows that the fastest
decaying matrix elements have decay rate proportional to $\lambda_1^2 N^2$. Furthermore, the model with a {\it non-collective, energy-conserving interaction},
where each qubit is coupled to an independent reservoir, can also
be solved explicitly \cite{PSE}. The fastest decay rate in this
case is shown to be proportional to $\lambda_1^2 N$.

2. As mentioned at the beginning of this section, we take the coupling constants $\lambda_1$, $\lambda_2$ so small that the resonances do not overlap (Condition (A1) in Section \ref{resultsection}). Consequently $\lambda_1^2 N^2$ and
$\lambda_2^2N$ are bounded above by $\Delta= 2\min_{j=1,
 \ldots,N} B_j$ (see also Remark 4 after Condition (A1)) and thus the decay
rates $\gamma_e$ do not increase indefinitely with
increasing $N$ in the regime considered here. Rather, the $\gamma_e$ are
attenuated by small coupling constants for large $N$. They are of
the order $\gamma_e\sim \Delta$.  We have shown that modulo an overall, common ($N$-dependent)
prefactor, the decay rates originating from the energy conserving
and exchanging interactions differ by a factor $N$.

In this paper we prove the results only for sufficiently high
temperatures. The general case will be treated elsewhere.

3.\ The decay of off-diagonal matrix elements in the energy basis does not relate directly to measurements of entanglement, \cite{PR,S}. We plan on elucidating the interplay between entanglement and decay of matrix elements in a subsequent work.

\medskip

{\bf Literature.\ } Collective decoherence has been studied
extensively in the literature. Among the many theoretical,
numerical and experimental works we mention here only
\cite{AHWBK,AR,BKT,DG,FF,PSE,UMY}, which are closest to the present work. We are
not aware of any prior work giving explicit decoherence rates of a
register for not explicitly solvable models, and without making
master equation technique approximations.

\section{Results}
\label{resultsection}

As mentioned in the introduction, we assume that
\begin{itemize}
\item[{\bf (A1)}] We have $C_0(|\lambda_1|+|\lambda_2|)N<\Delta$ for some constant $C_0$ (depending only on $g_1$, $g_2$, $J_{ij}$ and $B_j$). Here, $\Delta:=\min\{e-e': e,e'\in\spec(L_\s), e\neq e'\}$ is the gap in the spectrum of $L_\s$.
\end{itemize}
We implement a dynamical theory of resonances in a setting of {\it
spectral deformation} (see Section \ref{sect3}). This leads to the
following regularity requirement which we assume to be fulfilled
throughout the paper.
\begin{itemize}
\item[{\bf (A2)}] The function form factors $g_1$, $g_2$ in
\fer{6} satisfy the following  condition. For $h=g_1$ or $h=g_2$,
$$
h_{\beta}(u,\sigma) := \sqrt{\frac{u}{1-\e^{-\beta u}}}\
|u|^{1/2}\left\{
\begin{array}{ll}
h(u,\sigma) & \mbox{if $u\geq 0$}\\
\e^{\i\phi}\overline h(-u,\sigma) & \mbox{if $u<0$}
\end{array}
\right.
$$
is such that $\omega\mapsto
h_\beta(u+\omega,\sigma)$ has an analytic continuation, as a
map ${\mathbb C}\rightarrow L^2({\mathbb R}\times S^2,\d
u\times\d\sigma)$, into $\{|\omega|<\tau\}$, for some $\tau>0$.
Here, $\phi$ is an arbitrary fixed phase.
\end{itemize}

\noindent {\bf Remarks.\ } 1. Typically, the gap $\Delta$ depends
on $N$.  We have $\|H_\s\|< CN^2$ and $\|L_\s\|< CN^2$, for some
constant $C$. Therefore, if the $2^N$ ($=\dim\h_\s\otimes\h_\s$)
eigenvalues of $L_\s$ are roughly simple and equally distributed,
then the gap $\Delta$ is of the order $N^22^{-N}$. In this case,
Condition (A1) implies that the coupling constants $\lambda_1$ and
$\lambda_2$ have to be exponentially small in the size $N$ of the
qubit register. However, the gap $\Delta$ tends to become larger
as the multiplicities of the eigenvalues of $L_\s$ increase:
$\Delta$ is the minimal distance between {\it distinct}
eigenvalues of $L_\s$, spread over an interval of size $\|L_\s\|$.
Due to the increase of multiplicities, the gap may become
independent of $N$, as it happens in the following examples.
\\
\indent
 -- For $H_\s=J\sum_{j=1}^N S_j^zS_{j+1}^z$ (nearest
neighbour interaction; and say $S_{N+1}^z\equiv S_1^z$), we have
${\spec}(L_\s) = J \{-2N,-2N+1,\ldots, 2N-1, 2N\}$. It follows
that $\Delta=|J|$ is independent of $N$.

-- For $H_\s=\sum_{j=1}^N B_j S_j^z$, the difference between two
eigenvalues of $L_\s$ is given by $e-e'=\sum_{j=1}^N
B_j(n_j-n'_j)$, where $n_j,n'_j\in\{-2,0,2\}$. Hence (for $B_j>
0$), $\Delta = 2\min_{j=1,\ldots,N} B_j$.

2. Examples of form factors satisfying (A2) are
$g(k)=h_1(\sigma)|k|^p \e^{-|k|^2}$, where $p=-1/2+n$,
$n=0,1,2,\ldots$, and $h_1(\sigma)=\e^{\i\phi}\overline
h_1(\sigma)$. They include the physically most important cases,
see also \cite{MSB1,PSE}. We point out that it is possible to
weaken condition (A2) considerably, at the expense of a
mathematically more involved treatment, as mentioned in
\cite{MSB1}. The phase $\phi$ has been introduced, and its
physical interpretation has been given, in \cite{FM}.

\subsection{Effective dynamics of $\s$}
\label{mres1}

The main result of this section is Theorem \ref{thm1}, in which we
describe the effective dynamics of $\s$ and identify the dominant
part.

The evolution of reduced density matrix elements is governed by
exponentials $\e^{\i t\varepsilon_e^{(s)}(\lambda_1,\lambda_2)}$,
where $\varepsilon_e^{(s)}(\lambda_1,\lambda_2)$ are resonance
energies, lying in the upper complex plane. The subindex $e$ is
the eigenvalue of $L_\s$ which the resonance branches out of:
$\varepsilon_e^{(s)}(0,0)=e$, and the index $s=1,\ldots,\nu(e)\leq
d(e)$ distinguishes different resonance energies associated to the
same $e$ ($d(e)$ is the degeneracy of $e$ as an eigenvalue of $L_\s$). Using perturbation theory (we employ the Feshbach
projection method (see Section \ref{sect3} and \cite{BFS,MSB1})),
one obtains \fer{2.2}.

 Let $\{\eta_e^{(s,r)}\}_{r=1}^R$ and
$\{\widetilde\eta_e^{(s,r)}\}_{r=1}^R$ be bases of the eigenspaces
of the level shift operator $\Lambda_e$ and its adjoint $\Lambda_e^*$ (see \fer{5.1} for the formal definition of $\Lambda_e$),
\begin{eqnarray}
\Lambda_e \eta_e^{(s,r)} &=& \delta_e^{(s)}\eta_e^{(s,r)},\quad
r=1,\ldots R,
\label{2.4}\\
\Lambda_e^* \widetilde\eta_e^{(s,r)} &=&
\overline{\delta_e^{(s)}}\widetilde\eta_e^{(s,r)},\quad r=1,\ldots
R,
\label{2.5}
\end{eqnarray}
where $R=R(e,s)$ is the geometric multiplicity of the eigenvalue
$\delta_e^{(s)}$ of $\Lambda_e$. We choose bases that are {\it
dual} to each other,\footnote{This is always possible, see
Proposition \ref{appprop2} in Appendix \ref{app1}.} meaning that
\begin{equation}
\scalprod{\eta_e^{(s,r)}}{\widetilde\eta_e^{(s,r')}} =\delta_{r,r'}.
\label{dual}
\end{equation}
We define the projection
\begin{equation}
q_e^{(s)} =\sum_{r=1}^R |\eta_e^{(s,r)}\rangle\langle
\widetilde\eta_e^{(s,r)}|,
\label{projq}
\end{equation}
acting on the eigenspace of $L_\s$ associated to $e$.\footnote{ This projection is the same for all choices of bases $\eta_e^{(s,r)}$ and $\widetilde \eta_e^{(s,r)}$ satisfying \fer{dual}, as is easily verified using Proposition \ref{appprop2} of Appendix \ref{app1}.}
\begin{theorem}[Dynamics of matrix elements]
\label{thm1}
Denote by $\beta$ the inverse temperature of $\rr$. There is a $\lambda_0>0$ such that if
$\max\{|\lambda_1|,|\lambda_2|\}<\lambda_0/\beta$  then the limit \fer{2.9} exists for all
$\usigma,\utau$, and we have for $t\geq 0$
\begin{eqnarray}
\lefteqn{
 [\rhobar_t]_{\usigma,\utau}
-\langle\!\langle[\rhobar_\infty]_{\usigma,\utau}
\rangle\!\rangle=} \label{2.8}\\
&& \sum_{\{e, s: \ \varepsilon_{e}^{(s)}\neq 0\}} \e^{\i
t\varepsilon_{e}^{(s)}} \left[ {\sum_{\usigma',\utau'}}^*
\scalprod{\varphi_{\utau', \usigma'
}}{q_e^{(s)}\varphi_{\utau,\usigma}}
[\rhobar_0]_{\usigma',\utau'} +R_1\right]
+R_2(t). \nonumber
\end{eqnarray}
The $*$ in the last sum indicates that we only sum over spin configurations $\usigma', \utau'$ such that $e(\usigma',\utau')=-e$. The remainders satisfy
\begin{equation}
|R_1|\leq CN^2(\lambda_1^2+\lambda_2^2)\qquad\mbox{and} \qquad  |R_2(t)|\leq
CN^2(\lambda^2_1+\lambda^2_2) \e^{-\omega' t},
 \label{2.10}
\end{equation} where $C$ is a constant, $N$ is the register size,
and where $\omega'$ satisfies
$2\max_{e,s}\{\IM\varepsilon_e^{(s)}\}<\omega'<\tau/2$, with
$\tau$ given in Condition (A2).
\end{theorem}

{\bf Remarks.\ } 1.\ Since $q_e^{(s)}$ is a projection with range in the eigenspace associated to the eigenvalue $e$ of $L_\s$, we have $q_e^{(s)}\varphi_{\utau,\usigma}=0$ unless $e(\usigma,\utau)=-e$ (see the scalar product in \fer{2.8}).

2.\ The condition
$\max\{|\lambda_1|,|\lambda_2|\}<\lambda_0/\beta$ stems from the
particular complex deformation we choose in this work
(translation). A mathematically more sophisticated treatment,
involving a combination of spectral translation and dilation, and
an iterative renormalization group analysis will yield the theorem
for small $\lambda_1$, $\lambda_2$, but with a temperature
independent upper bound (see also \cite{MMS,MMS2} and remarks in
\cite{MSB1}).

3.\ We mention again that in this work, we consider the regime of non-overlapping resonances, described by Condition (A1) at the beginning of Section \fer{resultsection}. This means that $\lambda_1,\lambda_2\sim 1/N$.

\subsection{Non-interacting qubit register in magnetic field}
\label{sect01}

We consider the qubit register Hamiltonian \fer{2} with $J_{ij}=0$
and $B_j>0$, with a coupling to the reservoir given by \fer{6}. In
this section we  determine the resonance eigenvectors $\eta_
e^{(s,r)}$, $\widetilde\eta_e^{(s,r)}$ explicitly, as well as the
resonance energies $\varepsilon_e^{(s)}$ to lowest order in the
interaction, see Theorem \ref{thm2}. Those quantities are the key
ingredients entering the dynamics which we describe in Theorem
\ref{cor2} below.

Let $\usigma,\utau$ be spin configurations of the form \fer{11}.
Then
\begin{equation}
\varphi_{\usigma,\utau} = \varphi_{\sigma_1\tau_1}\otimes \cdots
\otimes \varphi_{\sigma_N\tau_N}\quad \mbox{with}\quad
\varphi_{\sigma\tau}=\varphi_\sigma\otimes\varphi_\tau\in
\cx^2\otimes\cx^2
 \label{u1}
\end{equation}
is an eigenvector of $L_\s$ with eigenvalue $e(\usigma,\utau)
=\sum_{j}B_j(\sigma_j-\tau_j)$.  The genericness condition
\fer{3.1} implies that if $\varphi_{\usigma,\utau}$ and
$\varphi_{\usigma',\utau'}$ are eigenvectors associated to the
same eigenvalue, then $\sigma_j-\tau_j=\sigma'_j-\tau'_j$ for all
$j$. If $\sigma_j-\tau_j=\pm 2$ then $\sigma_j=\pm 1$ and
$\tau_j=\mp 1$ are determined uniquely, while if
$\sigma_j-\tau_j=0$, then there are two choices,
$\sigma_j=\tau_j=\pm 1$. Consequently, an orthonormal basis of
eigenvectors of $L_\s$ associated to a given eigenvalue $e$ can be
constructed as follows. Take any {\it one} eigenvector
$\varphi_{\usigma,\utau}$ associated to $e$ and adjoin all
linearly independent vectors $\varphi_{\usigma',\utau'}$ with the
property $\{\sigma_j-\tau_j=0\} \Leftrightarrow
\{\sigma'_j-\tau'_j=0\}$. Thus, with each eigenvalue $e$ we
associate the number
\begin{equation}
N_0(e) = \{\mbox{number of indices $j$ s.t.  $\sigma_j=\tau_j$ in
any $(\usigma,\utau)$ with $e(\usigma,\utau)=e$} \}, \label{3.2}
\end{equation}
and the degeneracy of the eigenvalue $e$ of $L_\s$ is
$d(e)=2^{N_0(e)}$. To each eigenvalue $e$ of $L_\s$ there
corresponds a unique sequence of $N_0(e)$ indices indicating the
locations $j$ at which $\sigma_j=\tau_j$ for all $\usigma$,
$\utau$ associated with $e$. In other words, given $e$ there is a
unique sequence $\{\mu_k\}_{k=1}^{N_0(e)}$,
\begin{equation}
1\leq \mu_1 <\mu_2 < \cdots <\mu_{N_0(e)}\leq N,
 \label{3.5}
\end{equation}
having the property that any eigenvector $\varphi_{\usigma,\utau}$
associated to $e$ satisfies
\begin{equation}
\sigma_j=\tau_j\qquad \Longleftrightarrow \qquad j\in\{\mu_k:\
k=1,\ldots,N_0(e)\}.
 \label{3.6}
\end{equation}

Given an energy difference $e$ \fer{e}, and a sequence $\vb=
(\varrho_j)_{j=1}^{N_0(e)}$, $\varrho_j\in\{+1,-1\}$, we set
\begin{equation}
\delta_e^{(\vb)} = \lambda_1^2 [x_1(e)+\i y_1(e)]+
\lambda_2^2\big[x_2(e)
 +\i y_2(e)\big] +\sum_{j=1}^{N_0(e)} z_j^{\varrho_j},
 \label{3.9}
\end{equation}
where
\begin{eqnarray}
x_1(e) &=& -e_0\, {\rm P.V.}\scalprod{g_1}{\omega^{-1}g_1} \sum_{\{j:\ \sigma_j=\tau_j\}} \sigma_j\label{x1}\\
y_1(e) &=&\frac{\pi e_0^2 }{2\beta}
\gamma_+, \label{y1}
 \\
 x_2(e) &=&-\sum_{\{j:\ \sigma_j\neq
\tau_j\}} \!\!\sigma_j \ {\rm P.V.\,}\int_{\r} u^2 {\cal G}_2(2u)
\coth(\beta|u|) \frac{1}{u-B_j} \d u \label{3.8}\\
 y_2(e) &=&2\pi
\sum_{\{j:\ \sigma_j\neq \tau_j\}} B^2_j {\cal G}_2(2B_j)
\coth(\beta B_j), \label{3.11.1}\\
  z_j^{\pm} &=& \frac{1}{2}\left[ \i b_j(c_j+1) \pm
\sqrt{-b_j^2(c_j+1)^2 +4a[a-\i b_j(c_j-1)]}\right], \label{zj}
\end{eqnarray}
with
\begin{equation}
a=-\lambda^2_1e_0 \,{\rm P.V.}\scalprod{g_1}{\omega^{-1} g_1},
\quad b_j=4\pi\lambda^2_2\frac{B_j^2{\cal G}_2(2B_j)}{\e^{2\beta
B_j}-1},\quad c_j=e^{2\beta B_j}, \label{parameters}
\end{equation}
and
\begin{equation}
e_0 = e_0(e)= \sum_{j=1}^N (\sigma_j-\tau_j),\quad {\cal G}_k(u)
=\int_{S^2} |g_k(|u|,\Sigma)|^2 \d \Sigma, \quad
\gamma_+=\lim_{u\rightarrow 0_+} u \,{\cal G}_1(u). \label{e0}
\end{equation}
The form factors $g_1$, $g_2$ (see \fer{6}) are represented in
spherical coordinates  in \fer{e0} and ${\rm P.V.}$ stands for
principal value. Note that $e_0$ is the same for all spin
configurations $\usigma,\utau$ associated to the same energy
$e=e(\usigma,\utau)$. This follows from the genericness of the
magnetic field, \fer{3.1}, see paragraph after \fer{u1}. We show
in Theorem \ref{thmlso} that ${\rm Im}z_j^\pm\geq 0$. Let us
define the vectors
\begin{eqnarray}
\eta_e^{(\vb)} &=& \varphi_{\sigma_1\tau_1}\otimes\cdots\otimes
\xi_{\mu_1}^{\varrho_1}\otimes\cdots\otimes
\xi_{\mu_{N_0(e)}}^{\varrho_{N_0(e)}} \otimes
\cdots\otimes\varphi_{\sigma_N\tau_N}, \label{3.7}\\
\widetilde \eta_e^{(\vb)} &=&
\varphi_{\sigma_1\tau_1}\otimes\cdots\otimes \widetilde
\xi_{\mu_1}^{\varrho_1}\otimes\cdots\otimes \widetilde
\xi_{\mu_{N_0(e)}}^{\,\varrho_{N_0(e)}} \otimes
\cdots\otimes\varphi_{\sigma_N\tau_N}, \label{3.81}
\end{eqnarray}
where the $\varphi_{\sigma_{\mu_j}\tau_{\mu_j}}$ at positions
$\mu_j$, $j=1,\ldots,N_0(e)$, are replaced by $\xi, \widetilde\xi\in \cx^2\otimes\cx^2$, given by
\begin{eqnarray}
\xi_j^{\pm} &=& \varphi_{++} +\left[ 1+\i\frac{z_j^{\pm}-a}{b_jc_j}
\right]\varphi_{--} \label{3.3}\\
 \widetilde \xi_j^\pm &=& \varkappa^\pm_j\left( \varphi_{++} +\left[ 1+\i\frac{z_j^{\pm}-a}{b_jc_j}
\right]^*\varphi_{--}\right),
 \label{3.4}
\end{eqnarray}
with normalization constant $\varkappa_j^\pm =[1+b_j^{-2}c_j^{-1}\{ (b_jc_j-{\rm Im}z_j^\pm)^2  +(a-{\rm Re}z_j^\pm)^2\}]^{-1}$.

\begin{theorem}[Resonance energies and states]
 \label{thm2}
Let $e$ be an energy difference \fer{e} and let $\Lambda_e$ be the
associated level shift operator. The vectors $\eta_e^{(\vb)}$ and
$\widetilde \eta_e^{(\vb)}$, \fer{3.7} and \fer{3.81}, are bases
of eigenvectors of $\Lambda_e$ and its adjoint $\Lambda_e^*$,
respectively, which are dual to each other (see also \fer{dual}).
The eigenvalues of $\Lambda_e$ and $\Lambda_e^*$ associated to
$\eta_e^{(\vb)}$ and $\widetilde \eta_e^{(\vb)}$ are given by
$\delta_e^{(\vb)}$, \fer{3.9}, and its complex conjugate,
respectively. Furthermore, we have $\varepsilon_e^{(\vb)}
=e+\delta_e^{(\vb)}+O(\lambda_1^4+\lambda_2^4)$.
\end{theorem}

\noindent {\bf Remark.\ } The largest value of $N_0$ is $N$, which
corresponds to $e=0$, so $d(0)=2^N$. Here, $\mu_k=k$, $k=1,\ldots,
N$. The smallest value of $N_0$ is $0$, which corresponds to
$e=\pm e_{\rm max}$, where $e_{\rm max}=2\sum_j B_j$ is the
largest eigenvalue of $L_\s$. Thus $e_{\rm max}$ is a simple
eigenvalue of $L_\s$. Here, no two $\sigma_j$, $\tau_j$ are equal,
so the sequence $\{\mu_k\}$ is ``empty''. We have
$N_0(e)=N_0(-e)$, so $d(e)=d(-e)$ for all eigenvalues $e$.

\medskip
\noindent

The following result examines the resonance energies and shows
expression \fer{3.22.1} for the life times.

\begin{theorem}[Fermi Golden Rule Condition and decoherence rates]
\label{mlemmaX}
Assume that the so-called Fermi Golden Rule condition is satisfied:
\begin{equation}
\lambda_2^2 y_0:=4\pi\lambda_2^2\min_{j=1,\ldots,N}\{B^2_j{\cal G}_2(2B
_j)\coth(\beta B_j)\} > 0.\label{fgr2}
\end{equation}
There is a $c>0$
s.t. if\, $|\lambda_1|, |\lambda_2|<c$, then the decoherence rates
are given by \fer{3.22.1}, with
\begin{equation}
y_{12}(e) = \sum_{\{j:\ \sigma_j=\tau_j\}} \min\left\{ {\rm Im}z_j^+,{\rm Im}z_j^-\right\}.
\label{y12}
\end{equation}
\end{theorem}
{\bf Remark.\ } It is shown in Theorem \ref{thmlso} that ${\rm Im}z_j^\pm>0$ provided $ab_j\neq 0$, and that if $a=0$, then $z_j^+=4\pi \i\lambda_2^2 B^2_j {\cal G}_2(2G_j)\coth(\beta B_j)$, $z_j^-=0$ and if $b_j=0$ then $z_j^\pm=\pm a$. If $ab_j=0$ for all $j$ then $y_{12}(e)=0$.

\medskip

Let us illustrate how Theorems \ref{thm1}, \ref{thm2} and \ref{mlemmaX} combine to give the detailed dynamics of the register. Suppose that $\lambda_2\neq 0$. It is clear that for generic values of the magnetic field, all $\delta_e^{(\vb)}$ are different for different $\vb$ (see \fer{3.9}). Thus all resonance energies $\varepsilon_e^{(\vb)}=e+\delta_e^{(\vb)}+O(\lambda^2_1+\lambda_2^2)$ are simple, for small enough $\lambda_1,\lambda_2$. In this situation we obtain the following result:

\begin{theorem}[Dominant dynamics]
\label{cor2}
Suppose $\lambda_2\neq 0$ and suppose that the magnetic field is generic so that all $\delta_e^{(\vb)}$, \fer{3.9}, are distinct. There is a constant $c$ s.t. if $|\lambda_1|+|\lambda_2|<c$, then we have for all $\usigma$, $\utau$
\begin{eqnarray}
\lefteqn{ [\rhobar_t]_{\usigma,\utau} -\langle\!\langle
[\rhobar_\infty]_{\usigma,\utau}\rangle\!\rangle =} \label{3.15}\\
&& \sum_{\big\{ \vb:\
\varepsilon_e^{(\vb)}\neq 0\big\} } \e^{\i t\varepsilon_e^{(\vb)}} \left[   {\sum_{\usigma',\utau'}}^* w^{(e,\vb)}_{\usigma,\utau;\usigma',\utau'} \ [\rhobar_0]_{\usigma',\utau'}
+R_1\right] +R_2, \nonumber
\end{eqnarray}
where the $*$ means that we sum only over spin configurations s.t. $e(\usigma',\utau')=-e$, where  $\varepsilon_e^{(\vb)} = e+ \delta_e^{(\vb)}
+O(\lambda_1^4+\lambda_2^4)$, the remainder terms $R_1$, $R_2$ satisfy \fer{2.10}, and where
\begin{equation*}
w^{(e,\vb)}_{\usigma,\utau;\usigma',\utau'}=
\scalprod{\varphi_{\utau',\usigma'}}{\eta_e^{(\vb)}}
\scalprod{\widetilde \eta_e^{(\vb)}}{\varphi_{\utau,
\usigma}}\nonumber\\
=\prod_{j,k=1}^{N_0(e)} \scalprod{\varphi_{\tau'_{\mu_j},
\sigma'_{\mu_j}}}{\xi^{\varrho_j}_{\mu_j}}
\scalprod{\widetilde\xi_{\mu_k}^{
\varrho_k}}{\varphi_{\tau_{ \mu_k}, \sigma_{\mu_k}}}.
\end{equation*}
\end{theorem}

\subsection{Interacting qubit register in magnetic field}
\label{sect02}

In this section we consider the Hamiltonian $H_\s$, \fer{2}, with
{\it generic} parameters $J_{ij}$ and $B_j$. Energy differences of
$H_\s$ are
\begin{equation}
e(\usigma,\utau)=E(\usigma) - E(\utau) = \sum_{i,j=1}^N J_{ij}(\sigma_i\sigma_j -\tau_i\tau_j) +\sum_{j=1}^N B_j(\sigma_j-\tau_j).
\label{4.1}
\end{equation}
The condition $e(\usigma,\utau)=e(\usigma',\utau')$ is equivalent to
\begin{equation}
\sum_{i,j=1}^N J_{ij} m_{ij} + \sum_{j=1}^N B_j n_j=0,
\label{4.2}
\end{equation}
where $m_{ij}=\sigma_i\sigma_j-\sigma'_i\sigma'_j
-[\tau_i\tau_j-\tau'_i\tau'_j]$ and
$n_j=\sigma_j-\sigma'_j-[\tau_j-\tau'_j]$. For generic values of
$J_{ij}$ and $B_j$, the only solution of \fer{4.2} is $m_{ij}=0$,
$n_j=0$ for all $i,j=1,\ldots,N$.\footnote{Indeed, to solve
\fer{4.2} with some $m_{ij}$ or $n_j$ nonzero means to introduce
some correlations among the parameters $J_{ij}$ and $B_j$. Note
that in particular, $J_{ij}=J$ and $B_j=B$ is not a generic choice
of parameters.}

\begin{theorem}
\label{thm10} Let $e$ be a nonzero eigenvalue of $L_\s$. The
resonance energies associated to $e$ are
$\varepsilon_e(\usigma,\utau) =
e+\delta_e(\usigma,\utau)
+O(\lambda_1^4+\lambda_2^4)$, where $(\usigma,\utau)$ varies over
all spin configurations s.t. $e(\usigma,\utau)=e$, and where
(omitting $(\usigma,\utau)$ in the notation)
\begin{equation}
\delta_e = \lambda_1^2[ x_1 +\i y_1]+ \lambda_2^2 [x_2+\i y_2]
\label{4.3}
\end{equation}
with $x_1(e)$ and $y_1(e)$ given in \fer{x1} and \fer{y1}, and
\begin{eqnarray}
x_2 &=& -\frac12\sum_{k=1}^N {\rm P.V.}\int_{\r\times S^2} \!\!\! u^2 |g_2(u,\Sigma)|^2 \left[ \frac{|1-\e^{\beta u}|^{-1}}{u+v_k} + \frac{|1-\e^{-\beta u}|^{-1}}{u+v'_k} \right] \label{4.6} \\
y_2 &=& \frac{\pi}{2} \sum_{k=1}^N \left[ \frac{v_k^2 \, {\cal
G}_2(v_k)}{ |1-\e^{\beta v_k}|} + \frac{(v'_k)^2 \, {\cal
G}_2(v'_k)}{|1-\e^{-\beta v'_k}|} \right]. \label{4.7}
\end{eqnarray}
Here, $e_0$ is given in \fer{e0} and
\begin{equation}
v_k = -2\sigma_k \left[\sum_{j=1}^N (J_{jk}+J_{kj})\sigma_j +B_k\right], \quad v'_k = 2\tau_k \left[\sum_{j=1}^N (J_{jk}+J_{kj})\tau_j +B_k\right].
\label{4.8}
\end{equation}
The resonance eigenvectors associated to the resonance energy
$\varepsilon_e(\usigma,\utau)$ are
$\eta_{e(\usigma,\utau)}=\varphi_{\usigma,\utau}=\widetilde\eta_{e(\usigma,\utau)}$
(see \fer{2.4}, \fer{2.5}).
\end{theorem}
This result shows that the decoherence rates induced by the energy
conserving interaction are again maximally $O(\lambda_1^2N^2)$, as
in the case of the non-interacting register $(J_{ij}=0$). However,
the decoherence rates induced by the exchange interaction have a
complicated dependence on $N$: $y_2$ is a sum of $N$ terms each
one depending on $N$, the coupling parameters $J_{ij}$ and the
magnetic field $B_j$.

{\bf Remark.\ } One can proceed as for the non-interacting
register (Section \ref{sect01}) to analyze the resonances
bifurcating out of the origin (determined to lowest nontrivial
order by the spectrum of the level shift operator $\Lambda_0$).
One finds that $\Lambda_0$ has a simple eigenvalue at zero, and
that the imaginary part of the smallest (nonzero) resonance is
given by
\begin{equation}
\gamma_0 = 4\pi \lambda_2^2\min_{j=1,\ldots,N}\left\{
\frac{C_{j,+}^2 {\cal G}_2(2C_{j,+})}{|1-\e^{-2\beta C_{j,+}}|} +
\frac{C_{j,-}^2 {\cal G}_2(2C_{j,-})}{|1-\e^{-2\beta C_{j,-}}|}
\right\},
 \label{xxx1}
\end{equation}
where
$$
C_{j,\pm}=\sum_{k=1}^N(J_{jk}+J_{kj})\pm B_j.
$$

\section{Proofs}

\subsection{Proof of Theorem \ref{thm1}}
\label{sect3}

In Theorem \ref{prop4}, we first obtain a suitable expression for
the average $\av{A}_t$ of an observable $A\in B(\h_\s)$. This
result is based on the dynamical resonance theory developed in
\cite{MSB1}, see also \cite{JP1,JP2}, which we outline below.  In
a second step, we carry out a refined analysis of the resonance
theory to obtain Theorem \ref{prop5}. The combination of Theorems
\ref{prop4} and \ref{prop5} shows Theorem \ref{thm1}.

Let $A\in B(\h_\s)$. We have
\begin{eqnarray}
\langle A\rangle_t &=& {\rm Tr}_\s\left[\rhobar_t \ A \right] = {\rm Tr}_{\s+\rr}\left[ \rho_t \ A \otimes\bbbone_\rr\right]\nonumber\\
&=&\scalprod{\psi_0}{\e^{\i t\Ll}\left[
A\otimes\bbbone_\s\otimes \bbbone_\rr\right]\e^{-\i
t\Ll}\psi_0}. \label{2'}
\end{eqnarray}
In the last step, we pass to the {\it representation Hilbert
space} of the system (the GNS Hilbert space), where the initial
density matrix is represented by the vector $\psi_0$ (in
particular, the Hilbert space of the small system becomes
$\h_\s\otimes\h_\s$).

The dynamics of an observable $A$ is implemented by the group of
automorphisms $A\mapsto\e^{\i t\Ll} A \e^{-\i t\Ll}$. The
self-adjoint generator $\Ll$ is called the {\it Liouville
operator}. It is of the form $\Ll = L_0+\lambda_1 W_1+\lambda_2
W_2$, where $L_0=L_\s+L_\rr$ represents the uncoupled Liouville
operator,  and $\lambda_1 W_1+\lambda_2 W_2$ is the interaction
(represented in the GNS Hilbert space).

We take the initial state to be represented by the product  vector
$\psi_0=\psi_{\s,0}\otimes\psi_{\rr}$ (the product form of the
initial state is actually {\it not} necessary for our method to
work, see \cite{MSB1}). Here, $\psi_{\s,0}$ is an arbitrary
initial state of $\s$, and $\psi_{\rr}$ is the equilibrium state
of $\rr$ at a fixed inverse temperature $0<\beta<\infty$. We
denote by $\psi_{\s,\infty}$ the trace state of $\s$,
$\scalprod{\psi_{\s,\infty}}{(A_\s\otimes\bbbone_\s)\psi_{\s,\infty}}
= 2^{-N}{\rm Tr\,}(A_\s)$. We introduce the {\it reference vector}
\begin{equation}
\psi_{\rm ref} =\psi_{\s,\infty}\otimes\psi_\rr.
\label{refvect}
\end{equation}
The trace state has the separating property: given any state
$\psi_{\s,0}$ there is a (unique) operator $B\in {\frak M}_\s$,
satisfying $\psi_{\s,0}=(\bbbone_\s\otimes B)\psi_{\s,\infty}$. We
write $B':=\bbbone_\s\otimes B$ and note that $B'$ commutes with
all observables, so that we obtain from \fer{2'}
\begin{equation}
\langle A\rangle_t = \scalprod{\psi_0}{B' \e^{\i t\Ll}\left[
A\otimes\bbbone_\s\otimes \bbbone_\rr\right]\e^{-\i
t\Ll}\psi_{{\rm ref}}}.
\label{5.100}
\end{equation}
We now borrow a trick from the analysis of open systems far from
equilibrium: one can find a (non-self-adjoint) generator $\Kl$ s.t.
\begin{eqnarray*}
\e^{\i t\Ll} A \e^{-\i t\Ll} &=& \e^{\i t\Kl} A \e^{-\i t\Kl}
\mbox{\ \ for all observables $A$, $t\geq 0$, and}\\
\Kl\psi_{\rm ref} &=& 0.
\end{eqnarray*}
There is a standard way of constructing $\Kl$ given
$\Ll$ and the reference vector $\psi_{\rm ref}$. $\Kl$ is of
the form $\Kl =L_0+\lambda_1 I_1+\lambda_2 I_2$, where the interaction terms appearing in the expression for $\Ll$
undergo a modification $\lambda_1 W_1+\lambda_2 W_2 \rightarrow \lambda_1 I_1+\lambda_2 I_2$, c.f.
\cite{MSB1}. Formally, we may replace the
propagators in \fer{5.100} by those involving $\Kl$, and use that $\e^{-\i t\Kl}\psi_{\rm ref}=\psi_{\rm ref}$. This procedure has been carried out in a rigorous manner in \cite{MSB1}, yielding the following
resolvent representation
\begin{equation}
\langle A\rangle_t = -\frac{1}{2\pi \i}\int_{{\mathbb R}-\i}
\scalprod{\psi_0}{B'\,(\Kl(\omega)-z)^{-1}
\left[A\otimes\bbbone_\s\otimes\bbbone_\rr\right]\psi_{\rm ref}}
\e^{\i tz}\d z,
\label{eq20}
\end{equation}
where $\Kl(\omega)=L_0(\omega)+\lambda_1 I_1(\omega)+\lambda_2
I_2(\omega)$, $I_{1,2}$ are representing the interactions, and
$\omega\mapsto \Kl(\omega)$ is a spectral deformation
(translation) of $\Kl$. Relation \fer{eq20} holds for $0<{\rm Im} \omega<\tau$ (see condition (A2) in Section \ref{resultsection}); the integrand is analytic in that domain, and continuous as ${\rm Im}\omega\downarrow 0$. For $\omega\in\r$, the integrand is independent of $\omega$ and so it is constant for all $\omega$ in the domain of analyticity.

The spectral deformation is constructed as follows.
There is a deformation transformation $U(\omega)=\e^{-\i\omega
D}$, where $D$ is the (explicit) self-adjoint generator of
translations \cite{MSB1} transforming the operator $\Kl$ as
\begin{equation}
\Kl(\omega) = U(\omega) \Kl U(\omega)^{-1} = L_0+\omega N +\lambda_1 I_1(\omega) +\lambda_2 I_2(\omega).
\label{eq18}
\end{equation}
Here, $N$ is the total
number operator of $\h_\rr$, having spectrum ${\mathbb
N}\cup\{0\}$, where $0$ is a simple eigenvalue (vacuum eigenvector
$\psi_\rr$). For real values of $\omega$, $U(\omega)$ is a group
of unitaries. The spectrum of $\Kl(\omega)$ depends on ${\rm
Im\,}\omega$ and moves according to the value of ${\rm
Im\,}\omega$, whence the name ``spectral deformation''. Even
though $U(\omega)$ becomes unbounded for complex $\omega$, the
r.h.s. of \fer{eq18} is a well defined closed operator on a dense
domain, analytic in $\omega$ at zero. Analyticity is used in the
derivation of \fer{eq20} and this is where the analyticity
condition (A2) of Section \ref{resultsection} comes into play.

The point of the spectral deformation is that the (important part
of the) spectrum of $\Kl(\omega)$ is much easier to analyze
than that of $\Kl$, because the deformation uncovers the
resonances of $\Kl$. We have
$$
{\rm spec\,}\big(K_{0}(\omega)\big)=\{E_i-E_j\}_{i,j=1,\ldots,N}\bigcup_{n\geq 1}\{\omega n +{\mathbb R}\},
$$
because $K_0(\omega)=L_0+\omega N$, $L_0$ and $N$ commute, and the
eigenvectors of $L_0=L_\s+L_\rr$ are
$\varphi_i\otimes\varphi_j\otimes\psi_\rr$. The continuous
spectrum of $K_0$ is bounded away from the isolated eigenvalues by a gap of
size ${\rm Im\,}\omega$. The
operator $\lambda_1 I_1(\omega) +\lambda_2 I_2(\omega)$ is infinitesimally small with respect to the
number operator $N$, so for values of the coupling parameters
$\lambda_{1,2}$ small compared to ${\rm Im\,}\omega$, we can
follow the displacements of the eigenvalues by using analytic
perturbation theory. The following is an easy result (see e.g. \cite{MMS}).

\begin{theorem}
\label{thm3} Fix $\omega'$ s.t. $0<\omega'<\tau$
(where $\tau$ is as in Condition (A2) of Section \ref{resultsection}). There
is a constant $c_0>0$ s.t. if $\max\{|\lambda_1|,|\lambda_2|\}\leq
c_0/\beta$ ($\beta$ is the inverse temperature) then, for all $\omega$ with $\omega'<\omega<
\tau$,
the spectrum of $\Kl(\omega)$ in the complex half-plane $\{{\rm
Im\,}z<\omega'/2\}$ is independent of $\omega$ and consists purely
of the distinct eigenvalues
$$
\{\varepsilon_e^{(s)}:\  e\in{\rm spec}(L_\s), s=1,\ldots,\nu(e)\},
$$
where $1\leq\nu(e)\leq {\rm mult}(e)$ counts the splitting of the
eigenvalue $e$. Moreover, we have $\varepsilon_e^{(s)}(\lambda_1,\lambda_2)\rightarrow e$ as $\lambda_1,\lambda_2\rightarrow 0$, for all $e, s$, and furthermore,
${\rm Im\,}\varepsilon_e^{(s)}\geq 0$. Also, the continuous
spectrum of $\Kl(\omega)$ lies in the region $\{{\rm
Im\,}z\geq 3\omega'/4\}$.
\end{theorem}

Next we separate the contributions to the path integral in
\fer{eq20} coming from the singularities at the resonance energies
and from the continuous spectrum. We deform the path of
integration $z={\mathbb R}-\i$ into the line $z={\mathbb R}+\i
\omega'/2$, thereby picking up the residues of poles of the
integrand at $\varepsilon_e^{(s)}$ (all $e$, $s$). Let ${\mathcal
C}_e^{(s)}$ be a small circle around $\varepsilon_e^{(s)}$, not
enclosing or touching any other spectrum of $\Kl(\omega)$.
We introduce the (generally non-orthogonal) Riesz spectral
projections
\begin{equation}
Q_e^{(s)}  = Q_e^{(s)}(\omega,\lambda_1,\lambda_2) = -\frac{1}{2\pi\i}
\int_{{\mathcal C}_e^{(s)}} (\Kl(\omega)-z)^{-1} \d z.
\label{eq23}
\end{equation}
It follows from \fer{eq20} that
\begin{equation}
\langle A\rangle_t = \sum_e\sum_{s=1}^{\nu(e)} \e^{\i
t\varepsilon^{(s)}_e} \scalprod{\psi_0}{B'\, Q_e^{(s)}
[A\otimes\bbbone_\s\otimes\bbbone_\rr]\psi_{\rm ref}}
+R_2,
\label{eq21}
\end{equation}
where the remainder term $R_2$ comes from the contour integral enclosing the continuous spectrum and satisfies
\begin{equation}
|R_2|\leq C N^2 (\lambda_1^2+\lambda_2^2) \e^{-\omega' t/2},
\label{finrem}
\end{equation}
for some constant $C$ not depending on the dimension $N$ of
$\h_\s$, nor on $\lambda_1$, $\lambda_2$. Note that $R_2$ decays
faster in time than each term in the main part. The estimate
\fer{finrem} is a direct consequence of Proposition 4.2 in
\cite{MSB1} (see in particular equation (D.5) in the proof of this
proposition).

The ergodic mean time limits of $R_2$ and all terms in \fer{eq21} with
$\varepsilon_e^{(s)}\neq 0$ vanish, so
$$
\langle\!\langle A\rangle\!\rangle_\infty
:=\lim_{T\rightarrow\infty}\frac 1T \int_0^T \langle A\rangle_t \
\d t =\sum_{\{e, s:\ \varepsilon_e^{(s)} =0\}}\scalprod{\psi_0}{B'\,Q_0^{(s)}
[A\otimes\bbbone_\rr\otimes\bbbone_\rr]\psi_{\rm ref}}.
$$
Combining the latter expression with \fer{eq21} gives the following result.
\begin{theorem}
\label{prop4} For $\max\{|\lambda_1|,|\lambda_2|\}\leq c_0/\beta$ (see Theorem \ref{thm3}), we have for $t\geq 0$ and $A\in{\cal B}(\h_\s)$
\begin{equation}
\langle A\rangle_t -\langle\!\langle
A\rangle\!\rangle_\infty=\sum_{\{e,s:\
\varepsilon_e^{(s)}\neq 0\}} \e^{\i t\varepsilon_e^{(s)}}
\scalprod{\psi_0}{B'\, Q_e^{(s)}
[A\otimes\bbbone_\s\otimes\bbbone_\rr]\psi_{\rm ref}}
+R_2,
\label{eq21'}
\end{equation}
where $R_2$ satisfies \fer{finrem}.
\end{theorem}
Choosing $A = |\varphi_\utau\rangle\langle\varphi_\usigma|$ gives $\av{A}_t = [\rhobar_t]_{\usigma,\utau}$.
\begin{theorem}
\label{prop5}
Take $\max\{|\lambda_1|,|\lambda_2|\}\leq c_0/\beta$ (see Theorem \ref{thm3}), let $e$ be any eigenvalue of $L_\s$ and let $\usigma,\utau$ be spin
configurations. We have for $t\geq 0$
\begin{eqnarray*}
\lefteqn{ \scalprod{\psi_0}{B'\, Q_e^{(s)}
[|\varphi_\utau\rangle\langle\varphi_\usigma|
\otimes\bbbone_\s\otimes\bbbone_\rr]\psi_{\rm ref}} =}\nonumber\\
&&\sum_{\{ \usigma',\utau':\ e(\usigma',\utau')=-e \} }
\scalprod{\varphi_{\utau',\usigma'}}{q_e^{(s)}
\varphi_{\utau,\usigma}} [\rhobar_0]_{\usigma',\utau'}
+O\big(N^2(\lambda_1^2+\lambda_2^2)\big),
\end{eqnarray*}
where $q_e^{(s)}$ is defined in \fer{projq}, and $N=\dim\h_\s$ is the register size.
\end{theorem}
We point out that $q_e^{(s)}\varphi_{\utau,\usigma}$ vanishes
unless $e(\utau,\usigma)=e$. Theorem \ref{thm1} now follows
directly from Theorems \ref{prop4} and \ref{prop5}.

{\it Proof of Theorem \ref{prop5}.\ } Let $R$ be the rank of
$Q_e^{(s)}$ and $[Q_e^{(s)}]^*$, and let
$\{\chi_e^{(s,r)}\}_{r=1}^R$ and $\{\widetilde
\chi_e^{(s,r)}\}_{r=1}^R$ be bases of the ranges of those
projections {\it which are dual to each other}, so that, by Proposition \ref{linalgprop},
\begin{equation}
Q_e^{(s)} =\sum_{r=1}^R  |\chi_e^{(s,r)}\rangle \langle\widetilde\chi_e^{(s,r)}|.
\label{eq22'}
\end{equation}
We have
\begin{equation}
\Kl(\omega) \chi_e^{(s,r)} = \varepsilon_e^{(s)} \chi_e^{(s,r)} \quad \mbox{and} \quad
{}[\Kl(\omega)]^* \widetilde\chi_e^{(s,r)} =\overline{\varepsilon_e^{(s)}}\widetilde\chi_e^{(s,r)}.
\label{eq23'}
\end{equation}
The following isospectrality result is inferred from the Feshbach method, see \cite{BFS}, Section II and also \cite{MSB1}. We denote by $P_e$ the spectral projection of $K_0$ associated to the eigenvalue $e$, and we set $\Pbar_e=\bbbone-P_e$.
\begin{lemma}[Feshbach map]
\label{feshbachlemma}
Let $\chi$ be an eigenvector of $\Kl(\omega)$ with eigenvalue $\varepsilon$ (bifurcating out of $e$). Then $\xi=P_e\chi$ is an eigenvector of the operator
\begin{equation}
P_e\left[ e-I(\omega) \Pbar_e(\overline{K}_{\lambda_1,\lambda_2}(\omega) -\varepsilon)^{-1} \Pbar_e I(\omega)\right] P_e
\label{eq25'}
\end{equation}
with eigenvalue $\varepsilon$. Conversely, if $\xi\in{\rm Ran\,}P_e$ is an eigenvector of the operator \fer{eq25'} with eigenvalue $\varepsilon$, then
\begin{equation}
  \chi  =\left[ \bbbone - \Pbar_e (\overline{K}_{\lambda_1,\lambda_2}(\omega) -\varepsilon)^{-1}\Pbar_e I(\omega) \right] P_e\xi
\label{eq24'}
\end{equation}
is an eigenvector of $\Kl(\omega)$ with eigenvalue $\varepsilon$. Moreover, if $\chi$ is an eigenvector as above, then $\xi=P_e\chi\neq 0$ and conversely, if $\xi$ is an eigenvector as above, then $\chi$ given in \fer{eq24'} is nonzero. In particular, the geometric multiplicity of $\varepsilon$ as an eigenvalue of $\Kl(\omega)$ is the same as that of  $\varepsilon$ as an eigenvalue of \fer{eq25'}.
\end{lemma}
Expanding the resolvent in \fer{eq25'} around $(\Lbar_0(\omega)-e)^{-1}$  we obtain for $\xi=\xi_e^{(s,r)}$
\begin{equation}
\xi_e^{(s,r)} = \big[\eta_e^{(s,r)} +O\big(N^2(\lambda_1^2+\lambda_2^2)\big)\big]\otimes\psi_\rr,
\label{eq26'}
\end{equation}
where $\eta_e^{(s,r)}$ satisfies \fer{2.4}, with the level shift operator $\Lambda_e$ defined in \fer{5.1}, and where $N=\dim\h_\s$. We expand the resolvent in \fer{eq24'} around $(\Lbar_0(\omega)-e)^{-1}$ and use \fer{eq26'} to obtain
\begin{equation}
\chi_e^{(s,r)} = \left[ \bbbone - \Pbar_e(\Lbar_0(\omega)-e)^{-1}\Pbar_e I(\omega)\right] P_e \ \eta_e^{(s,r)}\otimes\psi_\rr +O\big(N^2(\lambda_1^2+\lambda_2^2)\big).
\label{eq27'}
\end{equation}
Proceeding in the same way we get the following representation for the eigenvectors $\widetilde\chi_e^{(s,r)}$ of the adjoint operator $[\Kl(\omega)]^*$,
\begin{equation}
\widetilde\chi_e^{(s,r)} = \left[ \bbbone - \Pbar_e(\Lbar_0(\overline\omega)-e)^{-1}\Pbar_e (I^*)(\overline\omega)\right] P_e \ \widetilde\eta_e^{(s,r)}\otimes\psi_\rr +O\big(N^2(\lambda_1^2+\lambda_2^2)\big),
\label{eq28'}
\end{equation}
where $\widetilde\eta_e^{(s,r)}$ satisfies \fer{2.5}. Relations \fer{eq27'} and \fer{eq28'} give
\begin{equation}
\chi_e^{(s,r)} = \eta_e^{(s,r)}\otimes\psi_\rr +O\big( N(|\lambda_1| +|\lambda_2|)\big),
\quad
\widetilde \chi_e^{(s,r)} = \widetilde\eta_e^{(s,r)}\otimes\psi_\rr +O\big( N(|\lambda_1| +|\lambda_2|)\big),
\label{eq31'}
\end{equation}
with the additional properties $\{\bbbone\otimes\langle
\psi_\rr|\} \chi_e^{(s,r)} =\eta_e^{(s,r)}\otimes\psi_\rr
+O\big( N^2(\lambda_1^2 +\lambda_2^2)\big)$ and
$\{\bbbone\otimes\langle\psi_\rr|\} \widehat \chi_e^{(s,r)}
=\widetilde\eta_e^{(s,r)}\otimes\psi_\rr +O\big(N^2(\lambda_1^2
+\lambda_2^2)\big)$.

Relations \fer{eq22'} and \fer{eq31'} show that
\begin{eqnarray}
\lefteqn{
\scalprod{\psi_0}{B'\, Q_e^{(s)}
[A\otimes\bbbone_\s\otimes\bbbone_\rr]\psi_{\rm ref}} =}\label{eq42'}\\
&&\sum_{r=1}^R \scalprod{\psi_0}{B'\, \eta_e^{(s,r)}\otimes
\psi_\rr} \scalprod{\widetilde\eta_e^{(s,r)}\otimes\psi_\rr}{
[A\otimes\bbbone_\s\otimes\bbbone_\rr]\psi_{\rm ref}}
+O\big(N^2(\lambda_1^2+\lambda_2^2)\big).\nonumber
\end{eqnarray}
Let us take $A=|\varphi_\utau\rangle\langle\varphi_\usigma|$. Then
 we have (see also after \fer{refvect})
\begin{equation}
 \scalprod{\widetilde \eta_e^{(s,r)}}{[ A\otimes\bbbone_\s]
\psi_{\s,\infty}} =2^{-N/2}
\scalprod{\widetilde\eta_e^{(s,r)}}{\varphi_{\utau,\usigma}}.
 \label{eq40'}
\end{equation}
Next we insert a decomposition of identity, and use again the
explicit form of $\psi_{\rm ref}$, \fer{refvect}, to obtain
\begin{eqnarray}
\scalprod{\psi_0}{B'\, \eta_e^{(s,r)}\otimes \psi_\rr} &=&
\sum_{\usigma',\utau'} \scalprod{\psi_0}{B'\,
\varphi_{\usigma',\utau'}\otimes\psi_\rr}
\scalprod{\varphi_{\usigma',\utau'}}{\eta_e^{(s,r)}}\nonumber\\
&=&2^{N/2} \sum_{\usigma',\utau'} \scalprod{\psi_0}{B'\,
 [|\varphi_{\usigma'}\rangle\langle\varphi_{\utau'}|\otimes\bbbone_\s
 \otimes\bbbone_\rr] \psi_{\rm ref}}
\scalprod{\varphi_{\usigma',\utau'}}{\eta_e^{(s,r)}}\nonumber\\
&=&2^{N/2} \sum_{\usigma',\utau'}
\scalprod{\varphi_{\usigma',\utau'}}{\eta_e^{(s,r)}}
[\rhobar_0]_{\utau',\usigma'}.
  \label{eq41'}
\end{eqnarray}
In the last step above, we commute $B'$ to the right,
$\scalprod{\psi_0}{B'\,
 [|\varphi_{\usigma'}\rangle\langle\varphi_{\utau'}
 |\otimes\bbbone_\s
 \otimes\bbbone_\rr] \psi_{\rm ref}} =
 \scalprod{\psi_0}{[|\varphi_{\usigma'}
 \rangle\langle\varphi_{\utau'}|\otimes
\bbbone_\s\otimes\bbbone_\rr]\psi_0} =
[\rhobar_0]_{\utau',\usigma'}$. Equations \fer{eq42'}-\fer{eq41'}
demonstrate Theorem \ref{prop5}. This also concludes the proof
of Theorem \ref{thm1}. \hfill $\square$

\subsection{Proof of Theorem \ref{thm2}}
\label{sect4}

The level shift operator associated to an eigenvalue $e$ of $L_\s$
is defined as
\begin{equation}
\Lambda_e = -P_e I \Pbar_e (\Lbar_0-e+\i 0)^{-1} \Pbar_e I P_e,
\label{5.1}
\end{equation}
where $P_e$ is the spectral projection of $L_0$ associated to
$\{e\}$, $\Pbar_e=\bbbone- P_e$ and $\Lbar_0$ is the operator
$L_0$ restricted to the range of $\Pbar_e$.\footnote{Relative to \cite{MSB1}, this definition differs by a sign.}  The interaction
operator $I$ has the form
\begin{equation}
I = V -J\Delta^{1/2} V J\Delta^{1/2},
 \label{5.2}
\end{equation}
where $J$ and $\Delta$ are the modular conjugation and the modular
operator associated to the pair $({\frak M},\psi_{\rm ref})$,
where $\psi_{\rm ref}$ is the reference vector given in
\fer{refvect}, and $\frak M$ is the von Neumann algebra of
observables.
The operator $V$ is given by
\begin{equation}
V = \lambda_1 (v_1\otimes \bbbone_\s)\otimes \phi_\beta(g_1)
+\lambda_2 (v_2\otimes\bbbone_\s)\otimes\phi_\beta(g_2).
 \label{5.3}
\end{equation}
Here, $v_1$ and $v_2$ are the non-demolition and energy exchange
interaction operators, see \fer{6}, and
$\phi_\beta(g)=\frac{1}{\sqrt 2}(a_\beta^*(g)+a_\beta(g))$ are the
thermal field operators.

\begin{theorem}
\label{thmlso}
{\rm (1) } The level shift operator $\Lambda_e$ has the form
\begin{equation}
\Lambda_e = \i\lambda_1^2 y_1(e) +\lambda_2^2[x_2(e)+\i y_2(e)]
+\sum_{\{j:\ \sigma_j=\tau_j\}} M^j, \label{lso1}
\end{equation}
where $y_1(e)$, $x_2(e)$ and $y_2(e)$ are given in \fer{y1},
\fer{3.8} and \fer{3.11.1}, and where the operator $M^j$ is
understood to act non-trivially only on the two dimensional
subspace, spanned by $\{\varphi_{++},\varphi_{--}\}$, of the
$j$-th factor $\cx^2\otimes\cx^2$ in the Hilbert space \fer{17}.
It is represented by the matrix
\begin{equation}
M^j =\left[
\begin{array}{cc}
a+\i bc & -\i bc\\
-\i b & -a+\i b
\end{array}
\right],
\label{M1}
\end{equation}
where $a$, $b$ and $c$ are given in \fer{parameters}.

{\rm (2)} The eigenvalues \fer{zj} of $M^j$ satisfy ${\rm
Im}(z_j^\pm)\geq 0$, and they are strictly positive if $ab\neq 0$.
For $a=0$, we have $z_j^+=\i b(c+1)$, $z_j^-=0$ and for $b=0$, we
have $z_j^\pm =\pm a$. The eigenvectors are given by \fer{3.3}.
\end{theorem}
Since $M^j$ acts non-trivially on different factors of the Hilbert
space for different $j$, we immediately see that the eigenvalues
of $\Lambda_e$ are the $\delta_e^{(\vb)}$, \fer{3.9}. This proves
Theorem \ref{thm2}.

\medskip

{\bf Proof of Theorem \ref{thmlso}}
 The level shift operators \fer{5.1}, for
interaction operators $V=\lambda G\otimes\bbbone_\s\otimes
\phi_\beta(g)$ and reference states
$\psi_{\s,\beta}\otimes\psi_\rr$ (equilibrium state for the
uncoupled dynamics), have been calculated explicitly in
\cite{MSB1}, Proposition 5.1. An easy adaptation to the present
situation gives the following result.
\begin{proposition}
\label{prop1}
We have the decomposition $\Lambda_e = \lambda_1^2
\Lambda_{e,1}  +
 \lambda_2^2 \Lambda_{e,2}$, where
 $\Lambda_\#=\lim_{\epsilon\downarrow 0}\Lambda_\#(\epsilon)$,
 with
\begin{eqnarray}
-2 \Lambda_{e,1}(\epsilon)&=&
P_e(v_1^2\otimes\bbbone-\bbbone\otimes v_1^2) P_e
\scalprod{g_1}{\frac{\omega}{\omega^2+\epsilon^2}g_1}\nonumber\\
&& -P_e (v_1\otimes\bbbone-\bbbone\otimes v_1)^2 P_e
\scalprod{g_1}{\coth(\beta\omega/2)\frac{\i
\epsilon}{\omega^2+\epsilon^2}g_1},
 \label{5.5}
\end{eqnarray}
where $\omega\geq 0$ is the radial variable (spherical
coordinates), and
\begin{eqnarray}
\lefteqn{ -2\Lambda_{e,2}(\epsilon)}\nonumber\\
 &=& P_e
(v_2\otimes\bbbone) \int_{\r\times S^2} \frac{u^2
|g_2(|u|,\Sigma)|^2}{|1-\e^{-\beta u}|} (L_\s
-e+u+\i\epsilon)^{-1} (v_2\otimes\bbbone)
P_e  \label{5.6.1}\\
&&+ P_e(\bbbone\otimes v_2) \int_{\r\times S^2} \frac{u^2
|g_2(|u|,\Sigma)|^2}{|1-\e^{+\beta u}|} (L_\s
-e+u+\i\epsilon)^{-1} (\bbbone\otimes v_2) P_e\label{5.6.2}\\
&&- P_e(v_2\otimes \bbbone) \int_{\r\times S^2} \frac{u^2
|g_2(|u|,\Sigma)|^2}{|1-\e^{-\beta u}|} (L_\s
-e+u+\i\epsilon)^{-1} (\bbbone\otimes v_2) P_e\label{5.6.3}\\
&&- P_e(\bbbone\otimes v_2) \int_{\r\times S^2} \frac{u^2
|g_2(|u|,\Sigma)|^2}{|1-\e^{+\beta u}|} (L_\s
-e+u+\i\epsilon)^{-1} (v_2\otimes\bbbone) P_e.
 \label{5.6.4}
\end{eqnarray}
All integration measures are $\d u\d\Sigma$, where $\d u$ is the
Lebesgue measure, and $\d\Sigma$ is the uniform measure on $S^2$.
\end{proposition}
\noindent
{\bf Remark.\ } There are no "cross-terms" involving products of
$v_1$ and $v_2$ in the expression \fer{5.1}: Indeed, for
instance
$$
\lambda_1\lambda_2 P_e \big(v_1\otimes\bbbone_\s
\otimes\phi_\beta(g_1) \big) \Pbar_e(\Lbar_0-e+\i 0)^{-1}\Pbar_e
\big(v_2\otimes\bbbone_\s\otimes\phi_\beta(g_2)\big) P_e=0,
$$
 since $S_j^x$ (occurring in $v_2$) maps any eigenspace of
$L_\s$ into its orthogonal complement.

\begin{proposition}
 \label{prop2}
We have $\Lambda_{e,2} = x_2(e)+\i y_2(e)+\i\Gamma_e$, where
$x_2(e)$ and $y_2(e)$ are given in \fer{3.8} and \fer{3.11.1}, and
where
\begin{equation}
\Gamma_e = \sum_{\{ j:\ \sigma_j=\tau_j\} } \Gamma^j,
 \label{5.7}
 \end{equation}
with
\begin{equation}
\Gamma^j = 4\pi B^2_j {\cal G}_2(2B_j) \left(1 -F_j\, \e^{-2\beta
B_j(S_j^z\otimes\bbbone) }\right) \big|1-\e^{-2\beta
B_j(S_j^z\otimes\bbbone)}\big|^{-1}.
 \label{5.8}
\end{equation}
The operator $\Gamma^j$ is understood to act non-trivially only on
the two-dimensional subspace, spanned by
$\{\varphi_{++},\varphi_{--}\}$, of the $j$-th factor
$\cx^2\otimes\cx^2$ in the Hilbert space \fer{17}. The "flip
operator" $F_j$ is defined by $F_j\varphi_{\usigma,\utau} =
\varphi_{\sigma_1\tau_1}\otimes\cdots \otimes
\varphi_{(-\sigma_j)(-\tau_j)}\otimes\cdots
\otimes\varphi_{\sigma_N\tau_N}$.

In the orthonormal basis $\{\varphi_{++},\varphi_{--}\}$, (see
also \fer{12}) the operator $\Gamma^j$ has the form
\begin{equation}
\Gamma^j = 4\pi\frac{B^2_j {\cal G}_2(2B_j)}{\e^{2\beta B_j}-1}
\left[
\begin{array}{cc}
\e^{2\beta B_j} & -\e^{2\beta B_j} \\
-1 & 1
\end{array}
\right].
 \label{5.9}
\end{equation}
\end{proposition}

{\it Proof.\ } We leave out the tensor product symbols $\otimes$ when
no confusion should occur. Take a $\varphi_{\usigma,\utau}$ in the
range of $P_e$. It follows from
$$
(v_2\otimes \bbbone)\varphi_{\usigma,\utau} =
\sum_{j=1}^N\varphi_{\sigma_1\tau_1} \cdots
\varphi_{(-\sigma_j)\tau_j}\cdots \varphi_{\sigma_N\tau_N}
$$
that
\begin{eqnarray}
L_\s (v_2\otimes\bbbone)\varphi_{\usigma,\utau} &=& \sum_{j=1}^N
\left[ \sum_{k\neq j} B_k (\sigma_k-\tau_k) +B_j(-\sigma_j
-\tau_j)\right]\varphi_{\sigma_1\tau_1} \cdots
\varphi_{(-\sigma_j)\tau_j}\cdots
\varphi_{\sigma_N\tau_N}\nonumber\\
&=& \sum_{j=1}^N (e-2B_j\sigma_j) \ \varphi_{\sigma_1\tau_1} \cdots
\varphi_{(-\sigma_j)\tau_j}\cdots \varphi_{\sigma_N\tau_N}.
 \label{5.12}
\end{eqnarray}
We now apply $P_e (v_2\otimes\bbbone)=P_e\sum_{k=1}^N
(S_k^x\otimes\bbbone)$ to \fer{5.12}. The only contribution is
coming from terms where $k=j$ in the resulting double sum: indeed,
$(S^x_kS_j^x\otimes\bbbone)\varphi_{\usigma,\utau}$ is orthogonal
to the range of $P_e$ unless $k=j$. It follows that
\begin{equation}
\fer{5.6.1} \,\varphi_{\usigma,\utau} = \int_{\r\times S^2} \d u\d
\Sigma\  \frac{u^2|g_2(|u|,\Sigma)|^2}{|1-\e^{-\beta u}|}
\sum_{j=1}^N (-2B_j\sigma_j+u+\i\epsilon)^{-1}
\varphi_{\usigma,\utau}.
 \label{5.13}
\end{equation}
In particular, the operator \fer{5.6.1} is diagonal in the energy basis.
Proceeding in the same fashion one finds
\begin{equation}
L_\s(\bbbone \otimes v_2)\varphi_{\usigma,\utau}  = \sum_{j=1}^N
(e+2B_j\tau_j)
\varphi_{\sigma_1\tau_1}\cdots\varphi_{\sigma_j(-\tau_j)} \cdots
\varphi_{\sigma_N\tau_N}
 \label{5.14}
\end{equation}
and
\begin{equation}
\fer{5.6.2} \,\varphi_{\usigma,\utau} = \int_{\r\times S^2} \d u\d
\Sigma\  \frac{u^2|g_2(|u|,\Sigma)|^2}{|1-\e^{+\beta u}|}
\sum_{j=1}^N (2B_j\tau_j+u+\i\epsilon)^{-1}
\varphi_{\usigma,\utau}.
 \label{5.15}
\end{equation}
The operator \fer{5.6.2} is thus also diagonal in the energy
basis. Next we consider \fer{5.6.3}. We apply
$P_e(v_2\otimes\bbbone)=P_e\sum_{k=1}^N(S_k^x\otimes\bbbone)$ to
\fer{5.14}. The only non-vanishing contribution comes from $k=j$
in the resulting double sum and only for terms where
$\sigma_j-\tau_j=0$. We obtain
\begin{eqnarray}
\fer{5.6.3}\, \varphi_{\usigma,\utau}
 &=&
-\!\!\sum_{\{j:\ \sigma_j=\tau_j\}} \int_{\r\times S^2} \d u\d
\Sigma\ \frac{u^2|g_2(|u|,\Sigma)|^2}{|1-\e^{-\beta
u}|}\nonumber\\
&&\times (2B_j\sigma_j+u+i\epsilon)^{-1}
\varphi_{\sigma_1\tau_1}\cdots\varphi_{(-\sigma_j)(-\tau_j)}\cdots
\varphi_{\sigma_N\tau_N}.
 \label{5.16}
\end{eqnarray}
Note that $\sigma_j$ can be replaced by $\tau_j$ in \fer{5.16}.
A similar argument gives
\begin{eqnarray}
\fer{5.6.4}\, \varphi_{\usigma,\utau}
 &=&
-\!\!\sum_{\{j:\ \sigma_j=\tau_j\}} \int_{\r\times S^2} \d u\d
\Sigma\ \frac{u^2|g_2(|u|,\Sigma)|^2}{|1-\e^{+\beta
u}|}\nonumber\\
&&\times (-2B_j\sigma_j+u+\i\epsilon)^{-1}
\varphi_{\sigma_1\tau_1}\cdots\varphi_{(-\sigma_j)(-\tau_j)}\cdots
\varphi_{\sigma_N\tau_N}.
 \label{5.17}
\end{eqnarray}
The operators \fer{5.6.3} and \fer{5.6.4} are not diagonal in the energy basis.

Next we use \fer{5.6.1}-\fer{5.6.4} and $\lim_{\epsilon\rightarrow
0_+}(-\alpha +u+\i\epsilon)^{-1} = -\i\pi\delta(u-\alpha) +{\rm
P.V.}\ \frac{1}{u-\alpha}$ in \fer{5.13} and \fer{5.15}-\fer{5.17}
to arrive at
\begin{eqnarray}
\Lambda_{e,2}\varphi_{\usigma,\utau}
&=& [x_2(e)+\i y_2(e)] \varphi_{\usigma,\utau} \label{5.23}\\
&&+4\i\pi \sum_{\{j:\ \sigma_j=\tau_j\}} B_j^2\frac{{\cal
G}_2(2B_j)}{|1-\e^{-2\beta B_j\sigma_j}|}
\varphi_{\usigma,\utau}\label{5.24}\\
&& -4\i\pi \sum_{\{j:\ \sigma_j=\tau_j\}} B_j^2 \frac{{\cal
G}_2(2B_j)}{|1-\e^{+2\beta B_j\sigma_j}|}
\varphi_{\sigma_1\tau_1}\cdots
\varphi_{(-\sigma_j)(-\tau_j)}\cdots \varphi_{\sigma_N\tau_N},
\qquad
 \label{5.25}
\end{eqnarray}
We have $|1-\e^{2\beta B_j\sigma_j}|^{-1}=\frac{\e^{-2\beta
B_j\sigma_j}}{|1-\e^{-2\beta B_j\sigma_j}|}$, so \fer{5.24} plus
\fer{5.25} combine to
\begin{equation}
4\i\pi \sum_{\{j:\ \sigma_j=\tau_j\}} B_j^2 {\cal
G}_2(2B_j)\left\{ 1- F_j\e^{-2\beta B_j\sigma_j} \right\}
\frac{1}{|1-\e^{-2\beta B_j\sigma_j}|}\varphi_{\usigma,\utau}.
 \label{5.26}
\end{equation}
The form \fer{5.9} of $\Gamma^j$ in the basis $\{\varphi_{++},\varphi_{--}\}$ is immediately obtained from \fer{5.8}. This completes the proof of Proposition \ref{prop2}. \hfill
$\square$

The following result follows directly from \fer{5.5}.

\begin{proposition}
\label{proplso1}
We have
\begin{equation}
\Lambda_{e,1} =  \i y_1(e)  -e_0 \,{\rm P.V.}
\scalprod{g_1}{\omega^{-1}g_1}\sum_{\{j:\ \sigma_j=\tau_j\}}
\sigma_j, \label{llso}
\end{equation}
where $y_1$ and $e_0$ are given in \fer{y1} and \fer{e0}.
\end{proposition}
We obtain \fer{lso1} now by combining Propositions \ref{prop2} and
\ref{proplso1}. This shows point (1) of Theorem \ref{thmlso}.
Point (2) is verified easily by using the expression \fer{zj}.
\hfill $\square$

\subsection{Proof of Theorem \ref{mlemmaX}}
According to \fer{3.17.1} we have $\gamma_e=\min\big\{
\IM\,\varepsilon_e^{(\vb)}\!:\ \vb\in \{+1,-1\}^{N_0(e)} \mbox{\
s.t. $\varepsilon_e^{(\vb)}\neq 0$ }\big\}$, where
$e=e(\usigma,\utau)=\sum_{j=1}^N B_j(\sigma_j-\tau_j)$.

-- For $e\neq 0$ and $\lambda_1$, $\lambda_2$ satisfying (A1), we
have $e+\delta_e^{(\vb)}\neq 0$, and hence
$\varepsilon_e^{(\vb)}\neq 0$, if $|\lambda_1|,|\lambda_2|<c$, for
some $c>0$. The smallest imaginary part of $\delta_e^{(\vb)}$ (for
$e$ fixed) is $\lambda_1^2y_1(e) +\lambda_2^2 y_2(e)+y_{12}(e)$.

-- For $e=0$, we have $e+\delta_e^{(\vb)} = \sum_{\{j:\ \sigma_j=\tau_j\}} z^{\varrho_j}_j$. Indeed, $e=0$
forces $\sigma_j=\tau_j$ for all $j$, and $e_0=0$ (see \fer{e0}).
It follows that $a_j=0$ and so
\begin{equation}
z_j^+=4\pi\i \lambda_2^2 B_j^2 {\cal G}_2(2B_j)\coth(\beta
B_j)\quad \mbox{and}\quad  z_j^-=0.
 \end{equation}
The smallest imaginary part of $\delta_e^{(\vb)}$ is thus zero,
corresponding to $\varrho_j=-1$ for all
$j=1,\ldots,N_0(0)=N$.\footnote{It can be inferred from general
considerations that at least one eigenvalue of $\Lambda_0$ must be
zero. Indeed, since the generator $\Kl$ has been designed to
annihilate the reference state $\psi_{\rm
ref}=\psi_{\s,\infty}\otimes\psi_\rr$, it follows that
$\Lambda_0\psi_{\s,\infty}=0$ \cite{M}. Note that indeed, for $\varrho_j=-1$, all $j$, we have $\eta_0^{(\vb)}=\psi_{\s,\infty}$.} All other imaginary parts
are strictly larger than the gap given by \fer{fgr2}.

This shows formula \fer{3.22.1} and completes the proof of Theorem \ref{mlemmaX}. \hfill $\square$

\subsection{Proof of Theorem \ref{thm10}}

Let $\varphi_{\usigma,\utau}$ be an eigenvector of $L_\s$ associated to the eigenvalue $e(\usigma,\utau)$. Let $k=1,\ldots,N$ be a fixed index. The vector $(S^x_k\otimes S^x_k) \varphi_{\usigma,\utau}$ is again an eigenvector of $L_\s$ with eigenvalue $e(\usigma',\utau')$, where $(\sigma'_j,\tau'_j)= (\sigma_j,\tau_j)$ for all $j\neq k$, and $(\sigma'_k,\tau'_k) = (-\sigma_k,-\tau_k)$. We now show that $e(\usigma,\utau)\neq e(\usigma',\utau')$ unless $e(\usigma,\utau)=0$. Indeed, suppose that $e(\usigma,\utau)= e(\usigma',\utau')$. Then, due to the genericness of the parameters $J_{ij}$ and $B_j$ (see after \fer{4.2}), we have $n_k=0$, from which it follows that $\sigma_k=\tau_k$. Furthermore, since $m_{ik}=0$ for all $i$, we obtain $\sigma_i=\tau_i$ for $i=1,\ldots,N$. We conclude that for all $k$,
\begin{equation}
P_{e(\usigma,\utau)} (S^x_k\otimes S^x_k)\varphi_{\usigma,\utau} = 0,\quad \mbox{if \ $e(\usigma,\utau)\neq 0$},
\label{10.1}
\end{equation}
where $P_e$ is the spectral projection of $L_\s$ associated to $e$. One sees also easily that if $k\neq l$, then $P_{e(\usigma,\utau)}(S^x_k\otimes S^x_l)\varphi_{\usigma,\utau}=0$.
\begin{proposition}
\label{prop10}
The level shift operators $\Lambda_e$ with $e\neq 0$ are diagonal in the energy basis. Their eigenvalues are given by $\lambda_1^2[x_1+\i y_1]+ \lambda_2^2[x_2+\i y_2]$ (see \fer{x1}, \fer{y1} and \fer{4.6}, \fer{4.7}).
\end{proposition}
{\bf Proof.\ } The spectrum of $\Lambda_{e,2}$ (see Proposition \ref{prop1}) is obtained as in the proof of Proposition \ref{prop2}. Relations \fer{5.12} and \fer{5.14} are replaced by
\begin{eqnarray}
L_\s (v_2\otimes\bbbone)\varphi_{\usigma,\utau} &=& \sum_{j=1}^N \left[ e(\usigma,\utau) +v_j(\usigma,\utau)\right] \varphi_{\sigma_1\tau_1}\cdots\varphi_{(-\sigma_j)\tau_j}\cdots \varphi_{\sigma_N\tau_N}
\label{10.2}\\
L_\s (\bbbone \otimes v_2)\varphi_{\usigma,\utau} &=& \sum_{j=1}^N \left[ e(\usigma,\utau) +v'_j(\usigma,\utau)\right] \varphi_{\sigma_1\tau_1}\cdots\varphi_{\sigma_j(-\tau_j)}\cdots \varphi_{\sigma_N\tau_N},
\label{10.3}
\end{eqnarray}
where $v_j$, $v'_j$ are given in \fer{4.7}. The terms \fer{5.16} and \fer{5.17} vanish due to \fer{10.1}. It then follows easily that the spectrum of $\Lambda_{e,2}$ is $\lambda_2^2[x_2+\i y_2]$.

The operator $\Lambda_{e,1}$ is the same as in the case
$J_{ij}=0$, so we can again use Proposition \ref{proplso1}, and, together with  Proposition \ref{prop1}, this gives the result. \hfill $\square$

\appendix

\section{Dual bases, projections, resonance eigenvectors}
\label{app1}

\begin{proposition}
\label{linalgprop}
Let $Q$ be a finite-dimensional projection in a Hilbert space $\h$.
Given any basis $\{\chi_r\}$ of ${\rm Ran\,}Q$, there is a unique basis $\{\widetilde\chi_r\}$ of ${\rm Ran\, }Q^*$ satisfying the duality condition $\scalprod{\chi_r}{\widetilde\chi_{r'}}=\delta_{r,r'}$. For $\chi_r$, $\widetilde\chi_r$ obtained in this way, we have $Q = \sum_{r} |\chi_r\rangle \langle \widetilde\chi_r|$.

\end{proposition}
{\bf Proof.\ } Take any basis $\{\chi_r\}$ of ${\rm Ran\,}Q$ and let $\psi\in\h$ be arbitrary. We have $Q\psi =\sum_{r}\chi_r c_r(\psi)$, where $\psi\mapsto c_r(\psi)\in\cx$ is a linear functional. Consequently, for each $r$, there is a $\widetilde\chi_r\in\h$ such that $c_r(\psi) =\scalprod{\widetilde\chi_r}{\psi}$. Hence $Q = \sum_{r}|\chi_r\rangle\langle\widetilde\chi_r|$. The vector $\chi_{r'}$ is left invariant by $Q$, so it follows that $\scalprod{\widetilde\chi_r}{\chi_{r'}}=\delta_{r,r'}$.

We show that the $\widetilde\chi_r$ are a basis of ${\rm Ran\, }Q^*$. Firstly, we have $Q^*\widetilde\chi_r=\widetilde\chi_r$ since $Q^*=\sum_{r}|\widetilde\chi_r\rangle\langle\chi_r|$ and $\scalprod{\widetilde\chi_r}{\chi_{r'}}=\delta_{r,r'}$, so we only need to show linear independence. Let $z_r$ be scalars. If $\sum_{r}z_r\widetilde\chi_r=0$, then, by taking the inner product with $\chi_{r'}$, where $r'$ is arbitrary, and using that $\scalprod{\widetilde\chi_r}{\chi_{r'}}=\delta_{r,r'}$, we see that $z_{r'}=0$.

To show uniqueness of $\{\widetilde\chi_r\}$ for fixed
$\{\chi_r\}$, we suppose that $\{\widetilde \alpha_r\}$ is another
dual basis. Then $\widetilde \alpha_r= \sum_{r'}\mu_{r,r'}
\widetilde\chi_{r'}$ and by the duality condition,
$\scalprod{\chi_r}{\widetilde\alpha_{r'}} = \delta_{r,r'} =
\mu_{r,r'}$, so $\mu$ is the identity. \hfill $\square$

\begin{proposition}
\label{appprop2} There are bases $\{\eta_e^{(s,r)}\}_r$ and
$\{\widetilde\eta_e^{(s,r)}\}_r$ of the eigenspaces of $\Lambda_e$
and $(\Lambda_e)^*$ associated to the eigenvalue $\delta_e^{(s)}$
and its complex conjugate, satisfying the duality property
\fer{dual}. Those bases are unique in the following sense: any
other pair of such bases $\{\alpha_e^{(s,r)}\}_r$,
$\{\widetilde\alpha_e^{(s,r)}\}_r$ is given by $\alpha_e^{(s,r)}=
\sum_{r'} [A]_{r,r'}\eta_e^{(s,r')}$ and  $\widetilde
\alpha_e^{(s,r)}= \sum_{r'} [(A^{-1})^*]_{r,r'}\widetilde
\eta_e^{(s,r')}$, where $A$ is an invertible matrix.
\end{proposition}
{\bf Proof.\ } From Proposition \ref{linalgprop} we know that we
can find bases $\{\chi_e^{(s,r)}\}_r$ and $\{\widetilde
\chi_e^{(s,r)}\}_r$ of the eigenspaces of $\Kl(\omega)$ and its
adjoint, associated to the eigenvalue $\varepsilon_e^{(s)}$ and
its complex conjugate, respectively, so that $\scalprod{\chi_e^{(s,r)}}{\widetilde\chi_e^{(s,r')}} =\delta_{r r'}$. Expansions \fer{eq27'}, \fer{eq28'} show
that $\scalprod{\eta_e^{(s,r)}}{\widetilde\eta_e^{(s,r')}}
=\lim_{\lambda_1,\lambda_2\rightarrow 0}
\scalprod{\chi_e^{(s,r)}}{\widetilde\chi_e^{(s,r')}}=\delta_{r,r'}$.

We know from the proof of Proposition \ref{linalgprop} that there
is a unique dual basis for fixed $\{\eta_e^{(s,r)}\}$. Thus, any
pair of dual basis is gotten by a change of basis of one
particular such pair. Let $\alpha_e^{(s,r)}$ be obtained by a base
change matrix $A$ as in the proposition. There is a unique
associated dual basis $\widetilde\alpha_e^{(s,r)} =
\sum_{r'}[B]_{r,r'}\widetilde \eta_e^{(s,r')}$. It is easy to see
that $\scalprod{\alpha_e^{(s,r)}}{\widetilde
\alpha_e^{(s,r')}}=\delta_{r,r'}$ implies that
$B=(A^{-1})^*$.\hfill $\square$

\section{Operators $\Kl$ and $\Kl(\omega)$}
\label{formulas}

The purpose of this Appendix is to provide some details on explicit formulas of the operators $\Kl$ and $\Kl(\omega)$. For more detail, we refer to \cite{MSB1}.

Smoothed-out creation and annihilation operators are defined by
\begin{equation*}
a^*(g) =\int_{\r^3} g(k)a^*(k)\d^3k, \qquad a(g) =\int_{\r^3}
\overline{g(k)}a(k)\d^3k,
\end{equation*}
for $g=g(k)\in L^2(\r^3,\d^3k)$, and the field operator is given by
\begin{equation}
\phi(g) = \frac{1}{\sqrt 2}[a^*(g) +a(g)].
 \label{7}
\end{equation}
The so-called Araki--Woods representation gives the Hilbert space (GNS) representation of the infinitely extended Bose gas in thermal equilibrium \cite{AW, Miqg}.\footnote{In this paper, we directly work in a spatially unitarily equivalent representation of the original representation, see \cite{MSB1} for details.} The Hilbert space is given by the bosonic Fock space over the one-particle space $L^2(\r\times S^2,\d^3 k\times\d\Sigma)$,
\begin{equation}
{\cal F} = {\cal F}(L^2(\r\times S^2,\d^3 k\times\d\Sigma)).
\label{gluedhsp}
\end{equation}
The thermal annihilation operators are
\begin{equation}
a_\beta(f) = a\big(\sqrt{1+\mu_\beta(u)} \chi_+(u) u f(u,\sigma)\big) - a^*\big(\e^{\i\phi} \sqrt{\mu_\beta(-u)} \chi_-(u) u \overline f(-u,\sigma)\big),
\label{new13}
\end{equation}
where $\mu_\beta(u)=(\e^{\beta u}-1)^{-1}$, $\chi_\pm$ are the indicator functions of $\r_\pm$, and $\phi\in\r$ is an arbitrary phase. The $a^*_\beta(f)$ are obtained by taking the adjoint on the r.h.s. of \fer{new13}. It is easy to see that the CCR are satisfied. The thermal field operator \fer{7} is thus represented by
\begin{equation}
\phi_\beta(f) = \frac{1}{\sqrt 2}(a_\beta^*(f)+a_\beta(f)) = \frac{1}{\sqrt 2}(a^*(f_\beta)+a(f_\beta))=:\phi( f_\beta),
\label{14''}
\end{equation}
for $f\in L^2(\r^3)$, where $f_\beta$ is defined in (A2), and where the $\phi$ in the r.h.s. is the field operator in $\cal F$.
The equilibrium state is represented by the vacuum vector of $\cal F$,
$\Omega_{{\rm R},\beta}=\Omega$.  For a one-body operator $O$ acting on wave functions of the variables $(u,\sigma)$, we write
\begin{equation}
\d\Gamma(O) = \int_{{\mathbb R}\times S^2}  a^*(u,\sigma) O
a(u,\sigma)\, \d u\d\sigma. \label{xx}
\end{equation}
for the second quantization of $O$. The dynamics of the field is generated by
\begin{equation}
L_\rr = \d\Gamma(u),
\label{94}
\end{equation}
the second quantization of the operator of multiplication by $u$. We have $L_\r\Omega_{\r,\beta}=0$, and for $z\in\cx$, $\e^{z L_\r} \phi_\beta(f)\e^{-zL_\r} = 2^{-1/2}\big[ a_\beta\big( \e^{-\overline z u}f\big) + a_\beta^*\big(\e^{z u}f\big)\big]$,
which gives the dynamics for $z=\i t$.

The Liouville operator $\Ll$ acting on $(\cx^N\otimes\cx^N)\otimes{\cal F}$ is given by
\begin{eqnarray}
\Ll &=& L_0 +\lambda_1 W_1+\lambda_2 W_2, \label{nr1}\\
L_0 &=& L_\s+L_\r = H_\s\otimes\bbbone_\s -\bbbone_\s\otimes H_\s + \d\Gamma(u),\label{nr2}\\
W_k &=& \sum_{j=1}^N S_j^k\otimes\bbbone_\s\otimes \phi((g_k)_\beta), \ \ k=1,2,\label{nr3}
\end{eqnarray}
where we understand $S_j^1=S_j^z$ and $S_j^2=S_j^x$.

The deformation group $U(\omega)$ (see after \fer{eq20}) is the {\it translation group} $U(\omega) = \e^{-\i\omega \d\Gamma(\i\partial_u)}$,
and the spectrally deformed Liouville operator is
\begin{equation}
\Ll(\omega) = L_0+\omega N +\lambda_1 W(\omega) +\lambda_2 W_2(\omega),
\label{sdlop}
\end{equation}
where $N=\d\Gamma(\bbbone)$ is the number operator in $\cal F$, and where $W_k(\omega)=\e^{-\omega\d\Gamma(\partial_u)} W_k\e^{\omega\d\Gamma(\partial_u)}$ (see also \fer{aa5}).

{\it Definition of the operator $\Kl$.\ }  This operator can be expressed in terms of the non-interacting Liouville operator $L_0$, the interaction $\lambda_1 W_1+\lambda_2 W_2$, see \fer{nr1}-\fer{nr3}, and the {\it modular data} $J,\Delta$ associated to the vector $\psi_{\rm ref}$, \fer{refvect}, and the von Neumann algebra $\mm={\cal B}(\h_\s)\otimes\bbbone_\s\otimes\mm_\beta$, where $\mm_\beta$ is the Weyl algebra of the Bose field (see e.g. \cite{BR,MSB1}). $J$ is an anti-unitary operator and $\Delta$ is a self-adjoint non-negative operator. The defining properties of $J$ and $\Delta$ are $J\Delta^{1/2} M\Omega_{\beta,0} = M^*\Omega_{\beta,0}$, for any $M\in\mm$, where $M^*$ is the adjoint operator of $M$. The explicit expressions are (see also \cite{BR,MSB1,MMS2})
\begin{eqnarray}
J=J_\s\otimes J_\rr \quad &\mbox{and}& \quad
\Delta = \Delta_\s\otimes\Delta_\rr, \label{5.01}\\
\Delta_\s &=& \e^{-\beta L_\s},\label{5.1'}\\
\Delta_\rr &=& \e^{-\beta L_\rr},\label{5.2'}\\
J_\s\phi_l\otimes\phi_r &=& \cc\phi_r\otimes \cc\phi_l,\label{5.3''}\\
J_\rr\psi_n(u_1,\sigma_1,\ldots,u_n,\sigma_n) &=& \e^{\i n\phi}\overline\psi_n(-u_1,\sigma_1,\ldots,-u_n,\sigma_n),\label{5.4'}
\end{eqnarray}
where the action of the antilinear operator $\cc$ is to take the complex conjugate of vector coordinates in the basis $\{\varphi_j\}_{j=1}^N$ of $\h_\s$, and $\overline\psi_n$ is the complex conjugate of $\psi_n\in{\cal F}$.  Relation \fer{5.4'} shows that $J_\rr a^\#(f(u,\sigma)) J_\rr = a^\#(\e^{\i\phi}\overline f(-u,\sigma))$, for $f\in L^2(\r\times S^2)$.

The interaction operators $I_k$ in \fer{eq18} are given by $I_k=W_k-W'_k$,  where
\begin{equation}
W_k' =J\Delta^{1/2} W_k J\Delta^{1/2} = \bbbone_\s\otimes \sum_{j=1}^N S_j^k \otimes \frac{1}{\sqrt{2}} \left[ a^*\big((g_k)_\beta\big)+ a\big(\e^{-\beta u}(g_k)_\beta \big) \right].
\label{9.1}
\end{equation}
The spectrally deformed operator $\Kl(\omega)$ is obtained as follows. The transformation of creation and annihilation operators under $U(\omega)$, \fer{sdlop}, is
\begin{equation}
U(\omega) a^\#(f) U(\omega)^{-1} = a^\#(f(\cdot +\omega)),\ \ \ \omega\in\r,
\label{aa1}
\end{equation}
where $f(\cdot +\omega)$ is the shifted function $(u,\sigma) \mapsto f(u+\omega,\sigma)$. Relation \fer{aa1} can be written in the form $U(\omega)a^\#(f)U(\omega)^{-1} = a^\#(\e^{\omega\partial_u}f)$. In order to obtain an analytic extension of \fer{aa1} to complex $\omega$, we need to take the complex conjugate of $\omega$ in the argument of the annihilation operator (since the latter is anti-linear in its argument). We thus have $I_k(\omega) = W_k(\omega) -W_k'(\omega)$, with
\begin{eqnarray}
W_k(\omega) &=&  \sum_{j=1}^N S_j^k\otimes\bbbone_\s\otimes\frac{1}{\sqrt{2}}\big[ a^*((g_k)_\beta(\cdot+\omega)) + a((g_k)_\beta(\cdot+\overline\omega))\big],\label{aa5} \\
W_k'(\omega) &=& \bbbone_\s\otimes\sum_{j=1}^N S_j^k \otimes \frac{1}{\sqrt{2}} \left[ a^*((g_k)_\beta(\cdot +\omega)) + a\big(\e^{-\beta (u+\overline\omega)}(g_k)_\beta(\cdot+\overline\omega) \big)\right].\ \ \ \ \ \ \ \ \label{aa6}
\end{eqnarray}
Finally, we have $\Kl(\omega) = \Ll(\omega)+\lambda_1 I_1(\omega)+\lambda_2I_2(\omega)$.

\end{document}